\documentclass[11pt]{article}
\usepackage{graphicx}
\usepackage{amssymb}
\usepackage{amscd}
\usepackage{mathrsfs}
\usepackage{longtable,lscape}
\usepackage{amsthm}
\usepackage{amsfonts}
\usepackage{amsmath}
\usepackage{bbm}
\usepackage{float}
\usepackage{url}

\newcommand{\compl}{{\mathbb C}}
\newcommand{\real}{{\mathbb R}}
\newcommand{\captionfonts}{\footnotesize}
\makeatletter  
\long\def\@makecaption#1#2{%
  \vskip\abovecaptionskip
  \sbox\@tempboxa{{\captionfonts #1: #2}}%
  \ifdim \wd\@tempboxa >\hsize
    {\captionfonts #1: #2\par}
  \else
    \hbox to\hsize{\hfil\box\@tempboxa\hfil}%
  \fi
  \vskip\belowcaptionskip}
\makeatother 
\begin{document}
\title{Modeling Meaning Associated with Documental Entities: Introducing the Brussels Quantum Approach}

\author{Diederik Aerts$^{1,2}$, Massimiliano Sassoli de Bianchi$^{1,3}$, Sandro Sozzo$^4$ and Tomas Veloz$^{1,5,6}$\\
\normalsize\itshape
$^1$ Center Leo Apostel for Interdisciplinary Studies and $^2$ Department of Mathematics,\\
\normalsize\itshape
Brussels Free University, Krijgskundestraat 33, 1160 Brussels (Belgium)\\
\normalsize
E-Mail:  \url{diraerts@vub.ac.be}\\ 
\normalsize\itshape
$^3$ Laboratorio di Autoricerca di Base, via Cadepiano 18, 6917 Barbengo (Switzerland)\\
\normalsize
E-Mail:  \url{msassoli@vub.ac.be}\\
\normalsize\itshape
$^4$ School of Business and Research Centre IQSCS, University of Leicester,\\
\normalsize\itshape
University Road, LE1 7RH Leicester (United Kingdom) \\
\normalsize
E-Mail: \url{ss831@le.ac.uk} \\
\normalsize\itshape
$^5$ Universidad Andres Bello, Departamento Ciencias Biol\'ogicas,\\
\normalsize\itshape
Facultad Ciencias de la Vida, 8370146 Santiago (Chile) \\
\normalsize\itshape
$^6$ Instituto de Filosof\'ia y Ciencias de la Complejidad,\\
\normalsize\itshape
 Los Alerces 3024, \~Nu\~noa, Santiago (Chile)\\
\normalsize
E-Mail: \url{tveloz@gmail.com}
}

\date{}
\maketitle
\begin{abstract}
\noindent
We show that the Brussels operational-realistic approach to quantum physics and quantum cognition offers a fundamental strategy for modeling the meaning associated with collections of documental entities. To do so, we take the World Wide Web as a paradigmatic example and emphasize the importance of distinguishing the Web, made of printed documents, from a more abstract meaning entity, which we call the Quantum Web, or QWeb, where the former is considered to be the collection of traces that can be left by the latter, in specific measurements, similarly to how a non-spatial quantum entity, like an electron, can leave localized traces of impact on a detection screen. The double-slit experiment is extensively used to illustrate the rationale of the modeling, which is guided by how physicists constructed quantum theory to describe the behavior of the microscopic entities. We also emphasize that the superposition principle and the associated interference effects are not sufficient to model all experimental probabilistic data, like those obtained by counting the relative number of documents containing certain words and co-occurrences of words. For this, additional effects, like context effects, must also be taken into consideration.
\end{abstract}

\section{Introduction\label{Introduction}}

In his book about the geometry of Information Retrieval (IR), Rijsbergen writes in the prologue~\cite{rijsbergen2004}: 
\begin{quote}
``Well imagine the world in IR before keywords or index terms. A document, then, was not simply a set of words, it was much more: it was a set of ideas, a set of concepts, a story, etc., in other words a very abstract object. It is an accident of history that a representation of a document is so directly related to the text in it. If IR had started with documents that were images then such a dictionary kind of representation would not have arisen immediately. So let us begin by leaving the representation of a document unspecified. That does not mean that there will be none, it simply means it will not be defined in advance. [...] a document is a kind of fictive object. Strangely enough Schr\"odinger [...] in his conception of the state-vector for QM envisaged it in the same way. He thought of the state-vector as an object encapsulating all the possible results of potential measurements. Let me quote: `It ($\psi$-function) is now the means for predicting probability of measurement results. In it is embodied the momentarily attained sum of theoretically based future expectation, somewhat as laid down in a catalogue.' Thus a state-vector representing a document may be viewed the same way -- it is an object that encapsulates the answers to all possible queries.''
\end{quote}
In the present chapter, we adopt that part of Rijsbergen's perspective that emphasizes the importance of distinguishing a corpus of written documents, like the pages forming the World Wide Web, made of actual (printed or printable) webpages, from the meaning (conceptual) entity associated with it, which in the case of the Web we simply call it the `Quantum Web' (in short, the `QWeb'),  because its modeling requires the use of notions derived from quantum theory, as we are going to discuss. This requirement is not at all accidental, and we are going to consider this crucial aspect too. Indeed, a strong analogy was established between the operational-realistic description of a physical entity, interacting with a measurement apparatus, and the operational-realistic description of a conceptual entity, interacting with a mind-like cognitive entity (see \cite{asdbs2015} and the references therein). In that respect, in a recent interpretation of quantum theory the strange behavior of quantum micro-entities, like electrons and photons, is precisely explained as being due to the fact that their fundamental nature is conceptual, instead of objectual (see \cite{aerts-etal2018} and the references therein). Considering the success of the quantum formalism in modeling and explaining data collected in cognitive experiments with human participants, it is then natural to assume that a similar approach can be proposed, \emph{mutatis mutandis}, to capture the information content of large corpora of written documents, as is clear that such content is precisely what is revealed when human minds interact with said documents, in a cognitive way. 

What we will describe is of course relevant for Information Retrieval (IR), {\it i.e.}, that \cite{melucci2015}: ``complex of activities performed by a computer system so as to retrieve from a collection of documents all and only the documents which contain information relevant to the user's information need.'' Although the term ``information'' is customarily used in this ambit, it is clear that the retrieval is about \emph{relevant} information, that is, \emph{meaningful} information, so that, in the first place, IR is really about \emph{Meaning Retrieval}. More specifically, similarly to a quantum measurement, an IR process is an interrogative context where a user enters a so-called \emph{query} into the system. Indeed, on a pragmatic level, a query works as an interrogation, where the system is \emph{asked} to provide documents whose meaning is strongly connected to the meaning conveyed by the query, usually consisting of a word, or sequence of words. In fact, since a search engine does not provide just a single document as an outcome, but an entire collection of documents, if the numerical values that are calculated to obtain the ranking are considered to be a measure of the outcome probabilities of the different documents, the analogy consists in considering the action of a search engine to be similar to that of an experimenter performing a large number of measurements, all with the same initial condition (specified by the query), then presenting the obtained results in an ordered way, according to their relative frequencies of appearance. Of course, the analogy is not perfect, as today search engines, when they look for the similarities between the words in the query and the documents, they only use deterministic processes in their evaluations. But we can certainly think of the deterministic functioning of today search engines as a provisional stage in the development of more advanced searching strategies, which in the future will also exploit non-deterministic processes, {\it i.e.}, probabilistic rankings (see for example \cite{Aerts2005}).

It is important to say, however, that our focus here is primarily on `the meaning that is associated with a collection of documents' and not on the exploration of more specific properties like `relevance' and `information need', which are more typically considered in IR. For the time being, our task is that of trying to find a way of modeling meaning content in a consistent way, and not yet that of considering the interplay between notions like `relevance' and `content', or `information need' and `user's request' \cite{melucci2015}. Our belief is that the adoption of a more fundamental approach, in the general modeling of meaning,  will help us in the future to also address in new and more effective ways those more specific properties and their relationships. 

Before entering in the description of our quantum approach, its motivations and foundations, it is useful to provide a definition of the terms ``meaning'' and ``concept,'' which we use extensively. By the term ``meaning,'' we usually refer to that content of a word, and more generally of any means of communication or expression, that can be conveyed in terms of concepts, notions, information, importance, values, etc. Meaning is also what different `meaning entities', like concepts, can share, and when this happens they become connected, and more precisely `connected through meaning'. By the term ``concept,'' we usually intend a well-defined and ideally formed thought, expressible and usable at different levels, like the intuitive, logical and practical ones. Concepts are therefore paradigmatic examples of `meaning entities', used as inputs or obtained as outputs of cognitive activities, for instance aimed at grasping and defining the essence of situations, decisions, reasoning, objects, physical entities, cultural artifacts, etc. Concepts are what minds (cognitive entities) are able to intend and understand, what they are sensitive to, and can respond to. They are what is created and discovered as the result of a cognitive activity, like study, meditation, observation, reasoning, etc. And more specifically, concepts are what minds use to make sense of their experiences of the world, allowing them, in particular, to classify situations, interpret them (particularly when they are new), connect them to previous or future ones, etc. 

An important aspect is that concepts, like physical entities, can be in different states. For instance, the concept \emph{Fruits},\footnote{We will generally indicate concepts using the italic style and the capitalization of the first letter, to distinguish them from the words used to designate them. So, we will distinguish the words ``juicy fruits,'' printed in a document, from the concept \emph{Juicy fruits}, which such words indicate. On the other hand, words written in italic style in the article but without capitalization of the first letter of the first word are just emphasized words.} when considered in the context of itself, can be said to be in a very neutral meaning-state, which is sometimes referred to as its `ground state'. But concepts can also be combined with other concepts, and when this is done their meaning change, {\it i.e.}, they enter into different states. For instance, the combination \emph{Sugary fruits} can be interpreted as the concept \emph{Fruits} in an excited state, because of the context provided by the \emph{Sugary} concept. But of course, it can also be interpreted as an excited state of the concept \emph{Sugary}, because of the context provided by the \emph{Fruits} concept.

An important notion when dealing with meaning entities like human concepts, is that of \emph{abstractness}, and its complementary notion of \emph{concretness}. For instance, certain concepts, like \emph{Table}, \emph{Chair} and \emph{House}, are considered to be relatively concrete, whereas other concepts, like \emph{Joy}, \emph{Entity} and \emph{Justice}, are considered to be relatively abstract. We can even order concepts in terms of their degree of concreteness, or abstractness. For example, the concept \emph{Table} can be considered to be more concrete than the concept \emph{Entity}; the concept \emph{Chess table} to be more concrete than the concept \emph{Table}; and the concept \emph{Alabaster chess table} to be more concrete than \emph{Chess table}, and so on. Here there is the idea that concepts are associated with a set of characteristic properties, and that by making their properties more specific, we can increase their degree of concreteness, up to the point that a concept possibly enters a one-to-one correspondence with an object of our spatiotemporal theater. This because, according to this view, concepts would typically have been created by abstracting them from objects. 

There is however another line to go from the abstract to the concrete, which can be considered to be more fundamental, and therefore also more important in view of a construction of a quantum model for the meaning content of a collection of documents. Indeed, although physical objects have played an important role in how we have formed our language, and in the distinction between abstract and concrete concepts, it is true that this line of going from the concrete to the abstract, linked to our historical need of naming the physical entities around us and define categories of objects having common features, remains a rather parochial one, in the sense that it does not take into full account how concepts behave in themselves, because of their non-objectual nature, particularly when they are combined, so giving rise to more complex entities having new emerging meanings. 

When this observation is taken into account, a second line of going from the abstract to the concrete appears, related to how we have learned to produce conceptual combinations to better think and communicate. The more abstract concepts are then those that can be expressed by single words, and an increase in concreteness is then the result of conceptual combinations, so that the most concrete concepts are those formed by very large aggregates of meaning-connected (entangled) single-word concepts, corresponding to what we would generically indicate as a \emph{story}, like those written in books, articles, webpages, etc. Of course, not a story only in the reductive sense of a novel, but in the more general sense of a cluster of concepts combined so as to create a well-defined meaning. It is this line of going from the abstract to the concrete that we believe is the truly fundamental, and in a sense also the universal one, which we will consider in our modeling strategy, when exploiting the analogy between a meaning retrieval situation, like when doing a Web search, and a quantum measurement in a physics' laboratory. But before doing this, in the next section we describe in some detail one of the most paradigmatic physics' experiments, which Feynman used to say that it contains the only mystery: the double-slit experiment.

In Section~\ref{interrogative}, we continue by providing a conceptualistic interpretation of the double-slit experiment, understanding it as an interrogative process. Then, in Section~\ref{modelWeb}, we  show how to use our analysis of the double-slit situation to provide a rationale for capturing the meaning content of a collection of documental entities. In Section~\ref{context}, we observe that quantum interference effects are insufficient to model all data, so that additional mechanisms, like context effects, need to be also considered. In Section~\ref{Conclusion}, we conclude our presentation by offering some final thoughts. In Appendix~\ref{context and interference}, we demonstrate that the combination of ``interference plus context effects'' allows in principle to model all possible data, while in Appendix~\ref{bond}, we introduce the notion of \emph{meaning bond} of a concept with respect to another concept, showing its relevance to the interpretation of our quantum formalism. 
\begin{figure}
\begin{center}
\includegraphics[width=14cm]{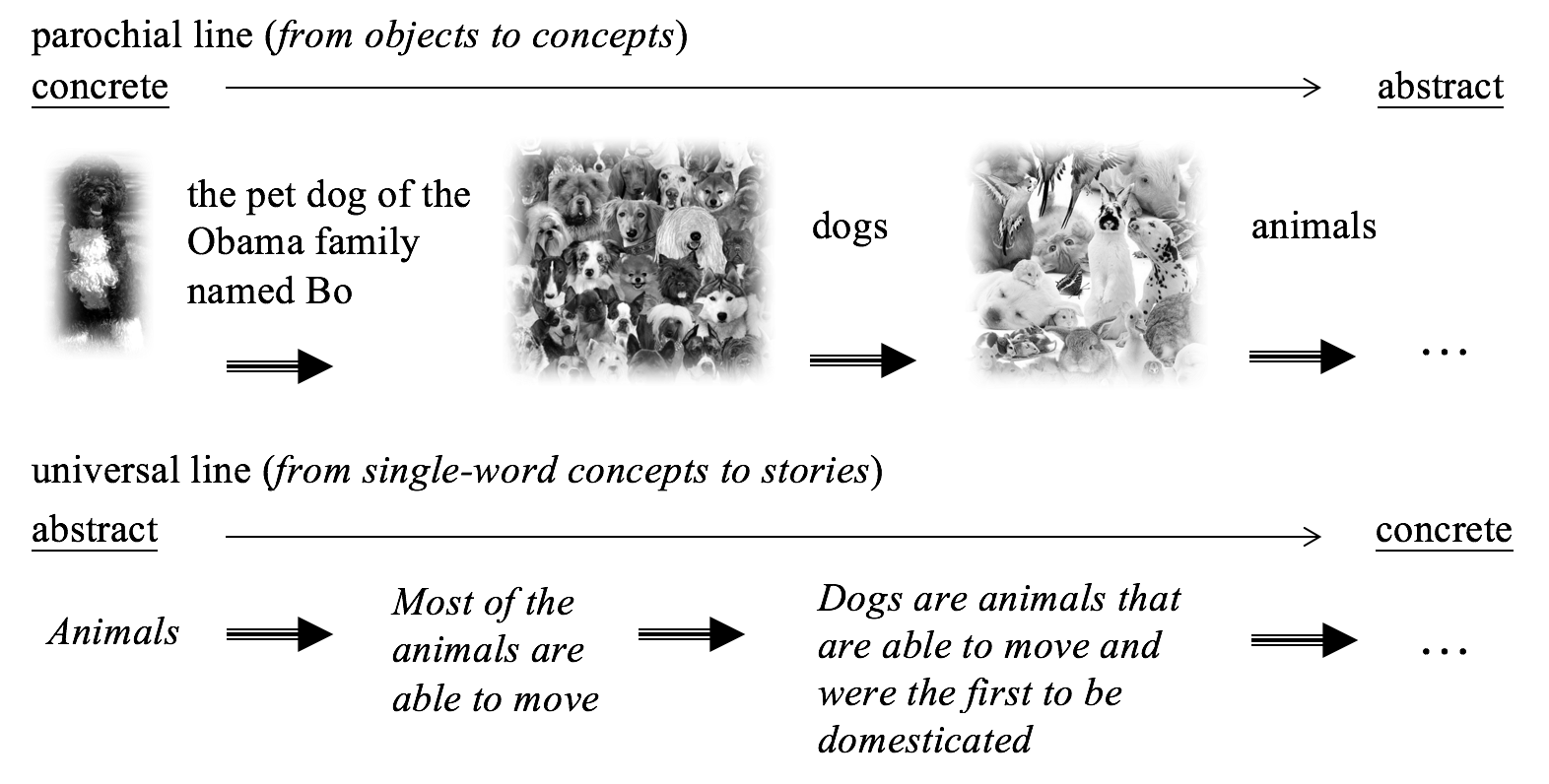}
\end{center}
\vspace{-15pt}
\caption{Two main lines connecting abstract to concrete exist in the human culture. The first one goes from concrete objects to more abstract collections of objects having common features. The second one goes from abstract single-word concepts to stories formed by the combination of many meaning-connected concepts.}
\label{Figure0}
\end{figure}

\section{The double-slit experiment\label{double-slit}}

The double-slit experiment is amongst the paradigmatic quantum experiments and can be used to effectively illustrate the rationale of our quantum modeling of the meaning content of corpora of written documents. One of the best descriptions of this experiment can be found in Feynman's celebrated lectures in physics \cite{Feynman1964}. We will provide three different descriptions of the experiment. The first one is just about what can be observed in the laboratory, showing that an interpretation in terms of particle or wave behaviors cannot be consistently maintained. The second (Sec.~\ref{interrogative}) one is about characterizing the experiment in a conceptualistic way, attaching to the quantum entities a conceptual-like nature, and to the measuring apparatus a cognitive-like nature. The third one is about interpreting the experiment as an IR-like process (Sec.~\ref{modelWeb}). 

\begin{figure}
\begin{center}
\includegraphics[width=15cm]{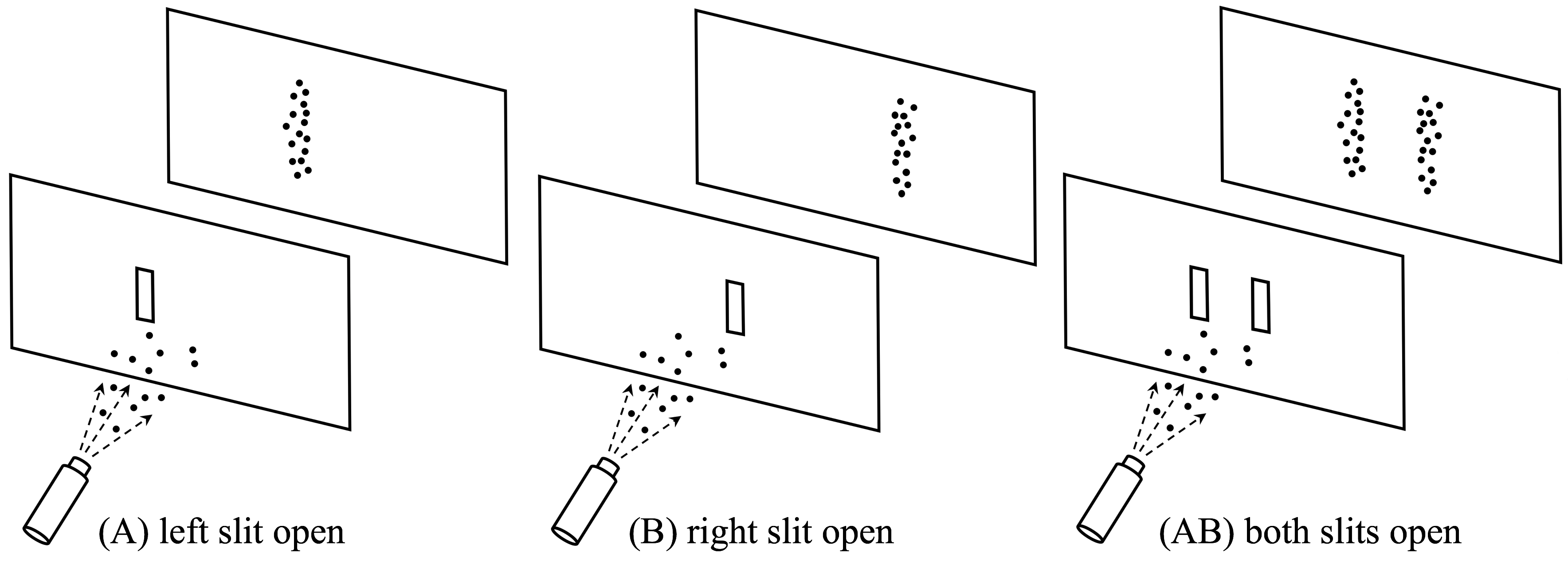}
\caption{A schematic description of the classical double-slit experiment, when: (A) only the left slit is open; (B) only the right slit is open and (AB) both slits are simultaneously open. Note that the time during which the machine gun fired the bullets in situation (AB) is twice than in situations (A) and (B).} 
\label{Figure1}
\end{center}
\end{figure}

We first consider the classical situation where the entities entering the apparatus, in its different configurations, are small bullets. Imagine a machine gun shooting a stream of these bullets over a fairly large angular spread. In front of it there is a barrier with two slits (that can be opened or closed), just about big enough to let a bullet through. Beyond the barrier, there is a screen stopping the bullets, absorbing them each time they hit it. Since when this happens a localized and visible trace of the impact is left on the screen, the latter functions as a detection instrument, measuring the position of the bullet at the moment of its absorption. Considering that the slits can be opened and closed, the experiment of shooting the bullet and observing the resulting impacts on the detection screen can be performed in four different configurations. The first one, not particularly interesting, is when both slits are closed. Then, there are no impacts on the detection screen, as no bullets can pass through the barrier. On the other hand, impacts on the detection screen will be observed if (A) the left slit is open and the right one is closed; (B) the right slit is open and the left one is closed; (AB) both slits are open. The distribution of impacts observed in these three configurations is schematically depicted in Figure~\ref{Figure1}.
As one would expect, the `both slits open' situation can be easily deduced from the two `only one slit open' situations, in the sense that if $\mu_A(x)$ and $\mu_B(x)$ are the probabilities of having an impact at location $x$ on the detection screen, when only the left (resp., the right) slit is open, then the probability $\mu_{AB}(x)$ of having an impact at that same location $x$, when both slits are kept open, is simply given by the uniform average: 
\begin{equation}
\mu_{AB}^{\rm \, bull}(x)={1\over 2}[\mu_A(x) + \mu_B(x)].
\label{classical}
\end{equation}

\begin{figure}
\begin{center}
\includegraphics[width=15cm]{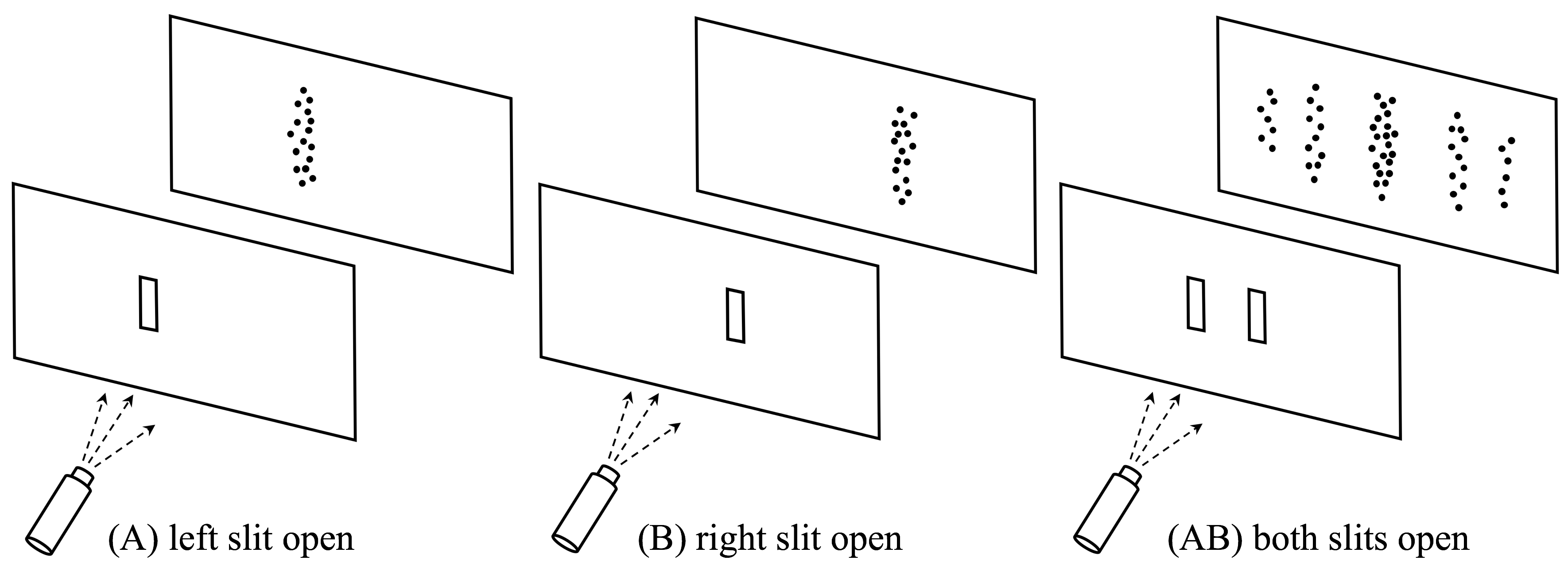}
\caption{A schematic description of the quantum double-slit experiment, when: (A) only the left slit is open; (B) only the right slit is open and (AB) both slits are simultaneously open. Different from the classical (corpuscular) situation, a fringe (interference) pattern appears when the left and right slits are both open.} 
\label{Figure2}
\end{center}
\end{figure}

Consider now a similar experiment, using electrons instead of small bullets. As well as for the bullets, well localized traces of impact are observed on the detection screen in the situations when only one slit is open at a time, always with the traces of impact distributed in positions that are in proximity of the open slit. On the other hand, as schematically depicted in Figure~\ref{Figure2}, when both slits are jointly open, what is obtained is not anymore deducible from the two `only one slit open' situations.
More precisely, when bullets are replaced by electrons, (\ref{classical}) is not anymore valid and we have instead:
\begin{equation}
\mu_{AB}^{\rm \, elec}(x)={1\over 2}[\mu_A(x) + \mu_B(x)]+{\rm Int}_{AB}(x),
\label{quantum}
\end{equation}
where ${\rm Int}_{AB}(x)$ is a so-called \emph{interference contribution}, 
which corrects the classical uniform average (\ref{classical}) and can take both positive and negative values.
Clearly, a corpuscular interpretation of the experiment becomes now impossible, as the region where most of the traces of impact are observed is exactly in between the two slits, where instead we would expect to have almost no impacts. Also, in the regions in front of the two slits, where we would expect to have the majority of impacts, practically no traces of impact are observed. 

\begin{figure}
\begin{center}
\includegraphics[width=7cm]{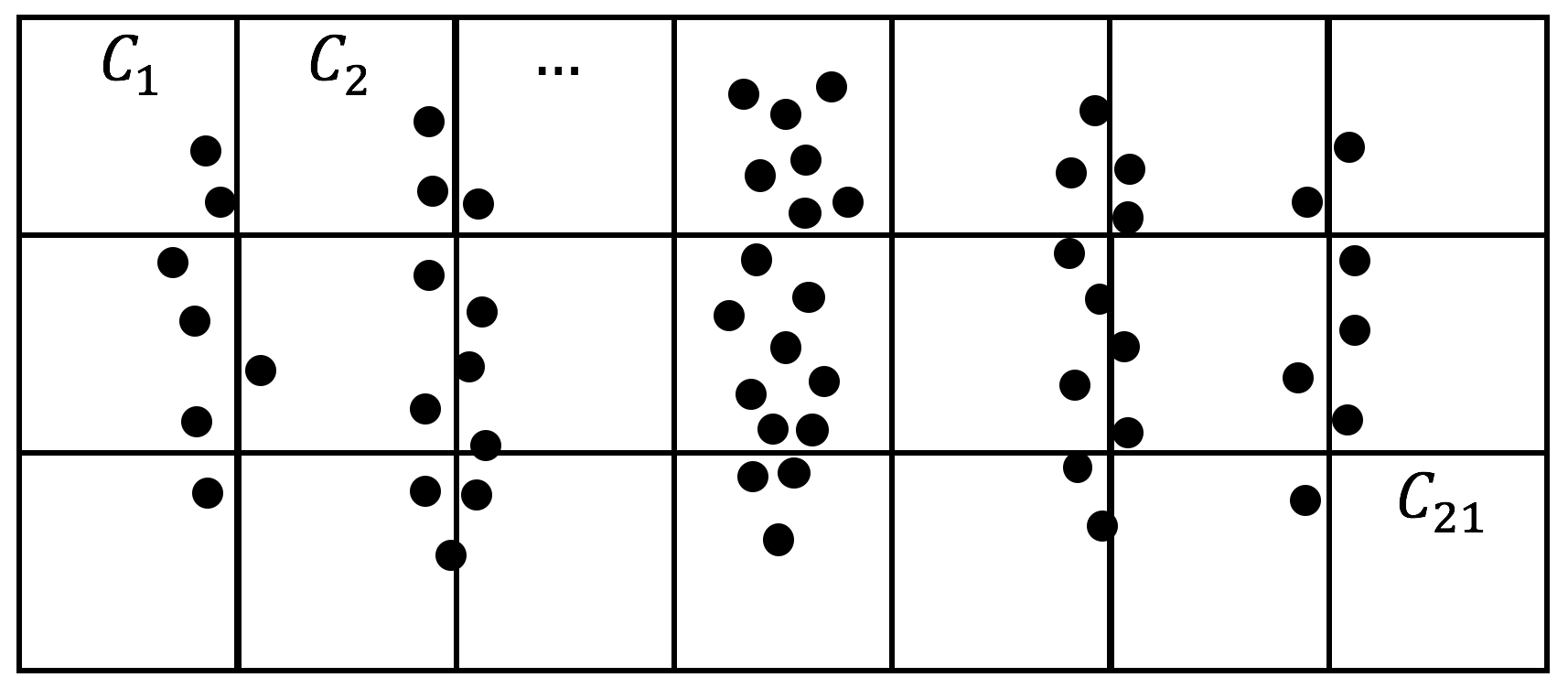}
\caption{The detection screen, partitioned into $n=21$ different cells, each one playing the role of an individual position detector, here showing the traces of $m=54$ impacts. The experimental probabilities are: $\mu_{AB}({\rm C}_1;21)={2\over 54}$, $\mu_{AB}({\rm C}_2;21)={2\over 54}$, $\mu_{AB}({\rm C}_3;21)={1\over 54}$, $\mu_{AB}({\rm C}_4;21)={7\over 54}$,\dots, $\mu_{AB}({\rm C}_{20};21)={1\over 54}$, $\mu_{AB}({\rm C}_{21};21)=0$.}
\label{Figure3}
\end{center}
\end{figure}

Imagine for a moment that we are only interested in modeling the data of the experiment (either with bullets or electrons) in a very instrumentalistic way, by limiting the description only to what can be observed at the level of the detection screen, {\it i.e.}, the traces that are left on it. For this, one can proceed as follows. The surface of the detection screen is first partitioned into a given number $n$ of numbered cells ${\rm C}_1,\dots,{\rm C}_n$ (see Figure~\ref{Figure3}). Then, the experiment is run until $m$ traces are obtained on it, $m$ being typically a large number. Also, the number of traces of impact in each cell is counted. If $m_{AB}({\rm C}_i)$ is the number of traces counted in cell ${\rm C}_i$, $i=1,\dots,n$, the experimental probability of having an impact in that cell is given by the ratio $\mu_{AB}({\rm C}_i;m)={m_{AB}({\rm C}_i)\over m}$. Similarly, we have $\mu_A({\rm C}_i;m)={m_{A}({\rm C}_i)\over m}$ and $\mu_B({\rm C}_i;m)={m_{B}({\rm C}_i)\over m}$, where $m_{A}({\rm C}_i)$ and $m_{B}({\rm C}_i)$ are the number of traces counted in cell ${\rm C}_i$ when only the left and right slits are kept open, respectively. If the experiments are performed using small bullets, one finds that the difference $\mu_{AB}({\rm C}_i;m)- {1\over 2}[\mu_A({\rm C}_i;m)+\mu_B({\rm C}_i;m)]$ tends to zero, as $m$ tends to infinity, for all $i=1,\dots,n$, whereas if the experiment is done using micro-entities, like electrons, it does not converge to zero, but towards a function ${\rm Int}({\rm C}_i)$, expressing the amount of deviation from the uniform average situation.

Now, once the three real functions $\mu_A({\rm C}_i;m)$, $\mu_B({\rm C}_i;m)$ and $\mu_{AB}({\rm C}_i;m)$ have been obtained, and their $m\to\infty$ limit deduced, one could say to have successfully modeled the experimental data, in the three different configurations of the barrier. However, a physicist would not be satisfied with such a modeling. Why? Well, because it is not able to explain why $\mu_{AB}({\rm C}_i)=\lim_{m\to\infty}\mu_{AB}({\rm C}_i;m)$ cannot be deduced, as one would expect, from $\mu_{A}({\rm C}_i)=\lim_{m\to\infty}\mu_{A}({\rm C}_i;m)$ and $\mu_{B}({\rm C}_i)=\lim_{m\to\infty}\mu_{B}({\rm C}_i;m)$, and why $\mu_{AB}({\rm C}_i)$ possesses such a particular interference-like fringe structure. So, let us explain how the quantum explanation typically goes. For this, we will need to exit the two-dimensional plane of the detection screen and describe things at a much more abstract and fundamental level of our physical reality. 

As is well-known, even if our description extends from the two-dimensional plane of the detection screen to the three-dimensional theater containing the entire experimental apparatus, this will still be insufficient to explain how the interference pattern is obtained. Indeed, electrons cannot be modeled as spatial waves, as they leave well-localized traces of impact on a detection screen, and they cannot be modeled as particles, as they cannot be consistently associated with trajectories in space.\footnote{This statement remains correct even in the de Broglie-Bohm interpretation of quantum mechanics, as in the latter the trajectories of the micro quantum entities can only be defined at the price of introducing an additional non-spatial field, called the quantum potential.} They are truly ``something else,'' which needs to be addressed in more abstract terms. And this is precisely what the quantum formalism is able to do, when describing physical entities in terms of the abstract notions of \emph{states}, \emph{evolutions}, \emph{measurements}, \emph{properties} and \emph{probabilities}, not necessarily attributable to a description of a spatial (or spatiotemporal) kind. 

So, let $|\psi\rangle$ be the state of an electron\footnote{One should say, more precisely, that $|\psi_{AB}\rangle$ is a Hilbert-space vector representation of the electron state, as a same state can admit different representations, depending on the adopted mathematical formalism.} (at a given moment in time) after having interacted with the double-slit barrier, with both slits open (we use here Dirac's notation). We can consider that this vector state has two components: one corresponding to the electron being reflected back towards the source (assuming for simplicity that the barrier cannot absorb it), and the other one corresponding to the electron having succesfully passed through the barrier and reached the detection screen. Let then $P_C$ be the projection operator associated with the property of ``having been reflected back by the barrier,'' and $P_{AB}$ the projection operator associated with the property of ``having passed through the two slits.' For instance, $P_C$ could be chosen to be the projection onto the set of states localized in the half-space defined by the barrier and containing the source, whereas  $P_{AB}$ would project onto the set of states localized in the other half-space, containing the detection screen.\footnote{Intuitively, one can also think of $P_{AB}$ as the projection operator onto the set of states having their momentum oriented towards the detection screen. Of course, all these definitions are only meaningful if applied to asymptotic states, viewing the interaction of the electron with the barrier as a scattering process, with the barrier playing the role of the local scattering potential.} We thus have $P_C +P_{AB}=\mathbb{I}$, and we can define $|\psi_{AB}\rangle={P_{AB}|\psi\rangle\over \| P_{AB}|\psi\rangle\|}$, which is the state the electron is in after having passed through the barrier and reached the detection screen region. Note that the barrier acts as a filter, in the sense that if the electron does leave a trace on the detection screen, we know it did succesfully pass through the barrier, and therefore was in state $|\psi_{AB}\rangle$ when detected. 

Now, since by assumption the $n$ cells ${\rm C}_i$ of the detection screen work as distinct measuring apparatuses, and an electron cannot be simultaneously detected by two different cells, for all practical purposes we can associate them with $n$ orthonormal vectors $|e_i\rangle$, $\langle e_i|e_j\rangle = \delta_{ij}$, corresponding to the different possible outcome-states of the position-measurement performed by the screen. This means that we can consider $\{|e_1\rangle, \dots, |e_n\rangle\}$ to form a basis of the subspace of states having passed through the barrier, and since we are not interested in electrons not reaching the detection screen, we can consider such $n$-dimensional subspace to be the effective Hilbert space ${\cal H}$  of our quantum system, which for instance can be taken to be isomorphic to the vector space $\compl^n$ of all $n$-tuples of complex numbers. 

According to the Born rule, the probability for an electron in state $|\psi_{AB}\rangle\in {\cal H}$,  to be detected by  cell ${\rm C}_i$, is given by the square modulus of the amplitude $\langle e_i|\psi_{AB}\rangle$, that is: $\mu_{AB}({\rm C}_i)=|\langle e_i|\psi_{AB}\rangle|^2$, and if we assume that an electron that has passed through the barrier is necessarily absorbed by the screen (assuming for instance that the latter is large enough), we have $\sum_{i=1}^n \mu_{AB}({\rm C}_i)=1$. Introducing the orthogonal projection operators $P_i=|e_i\rangle\langle e_i|$, we can also write, equivalently: 
\begin{equation}
\mu_{AB}({\rm C}_i)=\|P_i |\psi_{AB}\rangle\|^2 = \langle \psi_{AB} |P_i^\dagger P_i|\psi_{AB}\rangle = \langle \psi_{AB} |P_i^2|\psi_{AB}\rangle = \langle \psi_{AB} |P_i|\psi_{AB}\rangle.
\end{equation} 
More generally, if $I$ is a given subset of $\{1,\dots,n\}$, we can define the projection operator $M=\sum_{i\in I} P_i$, onto the set of states localized in the subset of cells with indexes in $I$, and the probability of being detected in one of these cells is given by: 
\begin{equation}
\mu_{AB}(i\in I)=\langle \psi_{AB} |M|\psi_{AB}\rangle=\sum_{i\in I} \mu_{AB}({\rm C}_i).
\label{Born}
\end{equation}
As an example, consider the situation of Figure~\ref{Figure3}, where one can for instance define the following seven projectors $M_k=P_k+P_{k+7}+P_{k+14}$, $k=1,\dots,7$, describing the seven columns of the $3\times 7$ screen grid. In particular, we have: $\mu_{AB}(i\in \{4,11,18\}) = {7\over 54}+{8\over 54}+{3\over 54}={1\over 3}$, {\it i.e.}, the probability for a trace of impact to appear in the central vertical sector of the screen (the central fringe) is one third. 

The double-slit  experiment does not allow to determine if an electron that leaves a trace of impact on the detection screen has passed through the left slit or the right slit. This means that the properties ``passing through the left slit'' and  ``passing through the right slit'' remain potential properties during the experiment, {\it i.e.}, alternatives that are not resolved and therefore (as we are going to see) can give rise to interference effects \cite{Feynman1964}. Let however write $P_{AB}$ as the sum of two projectors:  $P_{AB}= P_A + P_B$, where $P_A$ corresponds to the property of ``passing through the left slit'' and $P_B$ to the property of ``passing through the right slit.'' Note that there is no unique way to define these properties, and the associated projections, as is clear that electrons are not corpuscles moving along spatial trajectories. A possibility here is to further partition the half-space defined by $P_{AB}$ into two sub half-spaces, one incorporating the  left slit, defined by $P_A$ and the other one incorporating the right slit, defined by $P_B$, so that $P_AP_B=P_BP_A=0$. For symmetry reasons, we can assume that the electron has no preferences regarding passing through the left or right slits (this will be the case if the source is placed symmetrically with respect to the two slits), so that $\|P_A|\psi_{AB}\rangle\|^2=\|P_B|\psi_{AB}\rangle\|^2={1\over 2}$. We can thus define the two orthogonal states $|\psi_A\rangle= \sqrt{2}\, P_A|\psi_{AB}\rangle$ and $|\psi_B\rangle= \sqrt{2}\, P_B|\psi_{AB}\rangle$, and write:
\begin{equation}
|\psi_{AB}\rangle= (P_A + P_B)|\psi_{AB}\rangle= {1\over\sqrt{2}}(|\psi_A\rangle + |\psi_B\rangle).
\label{superposition}
\end{equation} 

According to the above definitions, $|\psi_A\rangle$ and $|\psi_B\rangle$ can be interpreted as the states describing an electron passing through the left and right slit, respectively.\footnote{Note however that, as we mentioned already, it is not possible to unambiguously define the two projection operators   $P_A$ and $P_B$, for instance because of the well-known phenomenon of the spreading of the wave-packet. In other words, there are different ways to decompose $|\psi_{AB}\rangle$ as the superposition of two states that can be conventionally associated with the one-slit situations, as per (\ref{superposition}). 
} In other words, in accordance with the quantum mechanical \emph{superposition principle}, we have expressed the electron state in the double-slit situation as a (uniform) superposition of one-slit states. Inserting (\ref{superposition}) in (\ref{Born}), now omitting the argument in the brackets to simplify the notation, we thus obtain: 
\begin{eqnarray}
\mu_{AB}&=&\langle \psi_{AB} |M|\psi_{AB}\rangle = {1\over 2}(\langle \psi_A| + \langle\psi_B|)M(|\psi_A\rangle + |\psi_B\rangle)\nonumber\\
&=& {1\over 2}(\langle \psi_A |M|\psi_A\rangle + \langle \psi_B |M|\psi_B\rangle + \langle \psi_A |M|\psi_B\rangle +\langle \psi_B |M|\psi_A\rangle)\nonumber\\
&=& {1\over 2}(\mu_A + \mu_B)+\underbrace{\Re\, \langle \psi_A |M|\psi_B\rangle}_{{\rm Int}_{AB}},
\label{interference}
\end{eqnarray}
where ${\rm Int}_{AB}$ is the interference contribution, with the symbol $\Re$ denoting the real part of a complex number, and we have used $\langle \psi_B |M|\psi_A\rangle = \langle \psi_A |M|\psi_B\rangle^*$. So, when there are indistinguishable alternatives in an experiment, as is the case here, since we can only observe the traces of the impact in the detection screen, without being able to tell through which slit the electrons have passed, states are typically expressed as a superposition of the states describing these alternatives, and because of that a deviation from the classical probabilistic average (\ref{classical}) will be observed, explaining in particular why an interference-like fringe-like pattern can form.\footnote{Of course, to characterize in detail such pattern one should explicitly solve the Schr\"odinger equation, which however would go beyond the scope of the present text.}

\section{Interrogative processes\label{interrogative}}

We now want to provide a cognitivistic/conceptualistic interpretation of the double-slit experiment, describing it as an interrogative process \cite{aerts-etal2018,AertsSassoli2017}. It is of course well understood that measurements in physics' laboratories are like interrogations. Indeed, when we want to measure 
 a physical quantity on a given physical entity, we can always say that we have a question in mind, that is: ``What is the value of such physical quantity for the entity?'' By performing the corresponding measurement, we then obtain an answer to the question. More precisely, the outcome of the measurement becomes an input for our human mind, which attaches to it a specific meaning, and it is only when such mental process has been completed that we can say to have obtained an answer to the question that motivated the measurement. In other words, there is a cognitive process, performed by our human mind, and there is a physical process, which provides an input for it. 

All this is clear, however, we want to push things further and consider that a measurement can also be described, per se, as an interrogative process, independently of a human mind possibly taking knowledge of its outcome. In other words, we also consider the physical apparatus as a cognitive entity, which answers a question each time it interacts with a physical entity subjected to a measurement, here viewed as a conceptual entity carrying some kind of meaning. This means that two cognitive processes are typically  involved in a measurement, one at the level of the apparatus, and another one at the level of the mind of the scientist interacting with it. The latter is founded on human meaning, but not the former, which is the reason why we  have to make as humans  a considerable effort to understand what is going on. In that respect, we can say that the construction of the theoretical and conceptual edifice of quantum mechanics has been precisely our effort in the attempt to understand the non-human meaning that is exchanged in physical processes, for instance when an electron interacts with a detection screen in a double-slit experiment. 

We will not enter here into the details of this \emph{conceptuality interpretation} of quantum mechanics, and simply refer to the review article \cite{aerts-etal2018} and to the references cited therein; this not only for understanding the genesis of this interpretation, but also for appreciating why it possibly provides a deep insight into the nature of our physical word. In the following, we limit ourselves to describing the double-slit experiment in a cognitivistic way, as this will be useful when we 
transpose the approach to an IR-like ambit. So, we start from the hypothesis that the electrons emitted by the electron gun are `meaning entities', {\it i.e.}, entities behaving in a way that is similar to how human concepts behave. And we also consider the detection screen to be a `cognitive entity', {\it i.e.}, an entity sensitive to the meaning carried by the electrons and able to answer questions by means of a written (pointillistic) language of traces of impact on its surface. We  
are then challenged as humans to understand the meaning of this language, and more precisely to guess the query that is  answered each time, and then see if the collection of obtained answers is consistent with the logic of such query.

There are of course different equivalent ways to formulate the question answered by the screen detector's mind. A possible formulation of it is the following: ``What is a good example of a trace of impact left by an electron passing through the left slit or the right slit?'' This way of conceptualizing the question is of course very 
``human," being based on the prejudice that the electron would be an entity always having spatial properties, which is not the case (this depends on its state). But we can here understand the ``passing through'' concept as a way to express the fact that the probability of detecting the electron by the final screen is zero if both slits are closed. An alternative way of formulating the same question, avoiding the ``passing through'' concept could be: ``What is a good example of an effect produced by an electron interacting with the barrier having both the left and right slits open?'' However, we will use in our reasoning the previous formulation of the question, as more intuitive for our spatially biased human minds. What we want is to explain the emergence of the fringe pattern by understanding the process operated by the detection screen, when viewed as a cognitive entity answering the above question. 

The first thing to observe is that such process will be generally indeterministic. Indeed, when we say ``passing through a slit", this is not sufficient to specify a unique trajectory in space for an electron (when assumed to be like a spatial corpuscle). This means that, if the screen cognitive entity thinks of the electron as a corpuscle, there are many ways in which it can pass through a slit, so, it will have to select one among several possibilities, which is the reason why, every time the question is asked, the answer (the trace of the impact on the screen) can be different (and cannot be predicted in advance), even though the state of the electron is always the same. The same unpredictability will manifest if the screen cognitive entity does not think of the electron as a spatial entity, but as a more abstract (non-spatial) conceptual entity, which can only acquire spatial properties by interacting with it. Indeed, also in this case the actualization of spatial properties will be akin to a \emph{symmetry breaking} process, whose outcomes cannot be predicted in advance.

To understand how the cognitive process of the screen detector entity might work, let us first concentrate on the central fringe, which is the one exhibiting the higher density of traces of impact and which is located exactly in between the two slits. It is there that the ``screen mind'' is most likely to manifest an answer. To understand the reason of that, we observe that an impact in that region elicits a maximum doubt as regards the slit the electron would have taken to cross the barrier, or even that it would have necessarily passed through either the left or the right slit, in an exclusive manner. Thus, an impact in that region is a perfect exemplification of the concept ``an electron passing through the left slit or the right slit." Now, the two regions on the screen that are exactly opposite the two slits, they have instead a very low density of traces of impact, and again this can be understood by observing that an answer in the form of a trace of impact there would be a very bad exemplification of the concept ``an electron passing through the left slit or the right slit," as it would not make us doubt much about the slit taken by the electron. Moving from these two low-density regions, we will then be back in situations of doubt, although less perfect than that of the central fringe, so we will find again a density of traces of impact, but this time less important, and then again regions of low density will appear, and so on, explaining in this way the alternating fringe pattern observed in experiments \cite{aerts-etal2018,AertsSassoli2017}.

\section{Modeling the QWeb\label{modelWeb}}

Having analyzed the double-slit experiment, and its possible cognitivistic/conceptualistic interpretation, we are now ready to transpose its narrative to the modeling of the meaning entity associated with the Web, which we have called the QWeb. Our aim is to provide a rationale for capturing the full meaning content of a collection of documental entities, which in our case will be the webpages forming the Web, but of course all we are going to say also works for other corpora of documents. As we explained in Sec.~\ref{Introduction}, there is a universal line for going from abstract concepts to more concrete ones, which is the one going from concepts indicated by single words (or few words) to those that are complex combinations of large numbers of concepts, which in our spatiotemporal theater can manifest as full-fledged stories, and which in our case we are going to associate to the different pages of the Web. Assuming they would have been ordered, we denote them ${\rm W}_i$, $i=1,\dots,n$. The meaning content of the Web has of course been created by us humans, and each time we interact with the webpages, for instance when reading them, cognitive processes will be involved, which in turn can give rise to the creation of new webpages. However, we will not be interested here in the modeling of these human cognitive activities, as well as when we model an experiment conducted in a physics' laboratory we are generally not interested in also modeling the cognitive activity of the involved scientists. 

As mentioned in Sec.~\ref{Introduction}, we want to fully exploit the analogy between an IR process, viewed as an interrogation producing a webpage as an outcome, and a measurement, like the position measurement produced by the screen detector in a double-slit experiment, also viewed as being the result of an interrogative process. So, instead of the $n$ cells ${\rm C}_i$, $i=1,\dots,n$, partitioning the surface of the detection screen, we now have the $n$ webpages ${\rm W}_i$, $i=1,\dots,n$, partitioning the Web canvas. What we now measure is not an electron, but the QWeb meaning entity, which similarly to an electron we assume can be in different states and that can transition to different possible outcome states when submitted to measurements.
 We will limit ourselves to measurements having the webpages ${\rm W}_i$ as their outcomes. More precisely, webpages ${\rm W}_i$ will play the same role as the cells ${\rm C}_i$ of the detection screen in the double-slit experiment, in the sense that we do not distinguish in our measurements the internal structure of a webpage, in the same way that we do not distinguish the locations of the impacts inside a single cell. So, similarly to what we did in Sec.~\ref{double-slit}, we can associate each webpage with a state $|e_i\rangle$, $i=1,\dots,n$, so that $\{|e_1\rangle,\dots, |e_n\rangle\}$ will form a basis of the $n$-dimensional QWeb's Hilbert state space. 

Let us describe the kind of measurements we have in mind for the QWeb. We will call them `tell a story measurements', and they consist in having the QWeb, prepared in a given state, interacting with an entity sensitive to its meaning, having the $n$ webpages stored in its memory, as concrete stories, so that one of these Web's stories will be told at each run of these measurements, with a probability that depends on the QWeb's state. The typical example of this is that of a search engine having the $n$ webpages stored in its indexes, used to retrieve some meaningful information, with the QWeb initial state being an expression of the meaning contained in the retrieval query (here assuming that the search engine in question would be advanced enough to also use indeterministic processes, when delivering its outcomes). 

If the state of the QWeb is $|e_i\rangle$, associated with the webpage ${\rm W}_i$, then the `tell a story measurement' will by definition provide the latter as an outcome, with probability equal to one. But the states $|e_i\rangle$, associated with the stories written in the webpages ${\rm W}_i$, only correspond, as we said, to the more concrete states of the QWeb, according to the definition of concreteness given in Sec.~\ref{Introduction}, and therefore only represent the tip of the iceberg of the QWeb's state space, as it would be the case for the position states of an electron. Indeed, the QWeb's states, in general, can be written as a superposition of the webpages' basis states:
\begin{equation}
|\psi\rangle = \sum_{j=1}^n r_j e^{i\rho_j}|e_i\rangle,\quad r_j,\rho_j\in \real, \quad r_j\ge 0, \quad \sum_{j=1}^{n}r_j^{2}=1.
\label{general-state}
\end{equation}
We can right away point out an important difference between (\ref{general-state}) and what is usually done in IR approaches, like the so-called \emph{vector space models} (VSM), where the states that are generally written as a superposition of basis states are those associated with the index terms used in queries (see for instance \cite{rijsbergen2004}, page 5, and \cite{melucci2015}, page 19). Here it is exactly the other way around: the dimension of the state space is determined by the number of available documents, associated with the outcome-states of the `tell a story measurements', interpreted as stories, \emph{i.e.}, as the more concrete states of the QWeb entity subjected to measurements. This also means that (as we will explain in the following) the states associated with single terms will not necessarily be mutually orthogonal, \emph{i.e.}, will not generally form a basis. Of course, another important difference with respect to traditional IR approaches is that the latter are built upon real vector spaces, whereas our quantum modeling is intrinsically built upon complex vector spaces (Hilbert spaces), where linearity works directly at the level of the complex numbers and weights are only obtained from the square of their moduli. In other words, the complex numbers $r_j e^{i\rho_j}$, appearing in the expansion (\ref{general-state}), can be understood as generalized coefficients expressing a connection between the meaning carried by the QWeb in state $|\psi\rangle$, and the meaning ``sticking out'' from (the stories contained in) the webpages ${\rm W}_j$.\footnote{More precisely, the real positive number $r_j$ can receive a specific interpretation as quantum \emph{meaning bonds}; see Appendix~\ref{bond}.}

As a very simple example of initial state, we can consider a state $|\chi\rangle$ expressing a \emph{uniform meaning connection} towards all the Web stories: $|\chi\rangle={1\over \sqrt{n}} \sum_{j=1}^n e^{i\rho_j}|e_j\rangle$, 
so that the probability to obtain story ${\rm W}_i$, in a `tell a story measurement', when the QWeb is in such uniform state $|\chi\rangle$, is:
\begin{equation}
\mu({\rm W}_i)=\langle\chi |P_i|\chi \rangle = {1\over n}\sum_{j,k=1}^n e^{i(\rho_j-\rho_k)}\underbrace{\langle e_k|e_i\rangle}_{\delta_{ki}}\underbrace{\langle e_i|e_j\rangle}_{\delta_{ij}} = {1\over n}.
\label{uniform}
\end{equation}
As another simple example, we can consider the QWeb state $|\chi_I\rangle={1\over \sqrt{m}} \sum_{j\in I} e^{i\rho_j}|e_j\rangle$, which is uniform only locally, \emph{i.e.}, such that only a subset $I$ of $m$ webpages , with $m\leq n$, would have the same (non-zero) probability of being selected as an actual story, so that in this case $\mu_I({\rm W}_i)=\langle\chi_I|P_i|\chi_I \rangle = {1\over m}$, if $i\in I$, and zero otherwise. 

It is important to observe that we are here viewing the QWeb as a whole entity, when we speak of its states, although it is clearly also a composite entity, in the sense that it is a complex formed by the combination of multiple concepts. Take two concepts $A$ and $B$ (for example, $A=$~{\it Fruits} and $B=${\it Vegetables}). As individual conceptual entities, they are certainly part of the QWeb composite entity, and as such they can also be in different states, which we can also write as linear combinations of the webpages' basis states: 
\begin{equation}
|\psi_A\rangle=\sum_{j=1}^n a_je^{i\alpha_j}|e_j\rangle,\quad |\psi_B\rangle=\sum_{j=1}^n b_je^{i\beta_j}|e_j\rangle,
\label{psiApsiB}
\end{equation}
with $a_j,b_j,\alpha_j,\beta_j\in \real$, $a_j,b_j\ge 0$, and $\sum_{j=1}^{n}a_j^{2}=\sum_{j=1}^{n}b_j^{2}=1$. 
These states, however, will be considered to be also states of the QWeb entity as a whole, as they also belong to its $n$-dimensional Hilbert space. In other words, even if states are all considered to be here states of the QWeb entity, some of them will also be interpreted as describing more specific individual conceptual entities forming the QWeb. We thus consider that individual concepts forming the composite QWeb entity can be viewed as specific states of the latter. Of course, the quantum formalism also offers another way to model composite entities, by taking the tensor product of the Hilbert spaces of the sub-entities in question. This is also a possibility, when modeling conceptual combinations, which proved to be very useful in the quantum modeling of data from cognitive experiments, particularly in relation to the notion of entanglement (see \cite{spinwind01,spinwind02} and the references cited therein), but in the present analysis we focus more directly on the superposition principle (and the interference effects it subtends) as a mechanism for accounting for the emergence of meaning when concepts are considered in a combined way \cite{a2009a} (see however the discussion in the first part of Sec.~\ref{context}).

Since we are placing ourselves in the same paradigmatic situation of the double-slit experiment, we want to consider how the combination of two concepts $A$ and $B$ -- let us denote the combination $AB$ -- can manifest at the level of the Web stories, in the ambit of a `tell a story measurement'. Here we consider the notion of `combination of two concepts' in a very general way, in the sense that we do not specify how the combination of $A$ and $B$ is actually implemented, at the conceptual level. In human language, if $A$ is the concept {\it Fruits} and $B$ is the concept {\it Vegetables}, their combination can for instance be  {\it Fruits-vegetables}, {\it Fruits and vegetables}, {\it Fruits or vegetables}, {\it Fruits with vegetables}, {\it Fruits are sweeter than vegetables}, etc., 
which of course carry different meanings, {\it i.e.}, describe different states of their two-concept combination. In fact, also stories which are jointly about {\it Fruits} and {\it Vegetables} can be considered to be possible states of the combination of these two concepts. All these possibilities give rise of different states $|\psi_{AB}\rangle$, describing the combination of the two concepts $A$ and $B$. 

These two concepts can be seen to play the same role of the two slits in the double-slit experiment. When the two slits are jointly open, we are in the same situation as when the two concepts $A$ and $B$ are jointly considered in the combination $AB$, producing a state $|\psi_{AB}\rangle$ that we can describe as the superposition of two states $|\psi_A\rangle$ and $|\psi_B\rangle$, which are the states of the concepts $A$ and $B$, respectively, when considered not in a combination, and which play the same role as the states of the electron in the double-slit experiment traversing the barrier when only one of the two slits is kept open at a time. Of course, different superposition states can in principle be defined, each one describing a different state of the combination of the two concepts, but here we limit ourselves to the superposition (\ref{superposition}), where the states $|\psi_A\rangle$ and $|\psi_B\rangle$ have the exact same weight in the superposition.

Let now $X$ be a given concept. It can be a concept described by a single word, or a more complex concept, described by the combination of multiple concepts. We consider the projection operator $M_X^w$, onto the set of states that are \emph{manifest} stories about $X$. This means that we can write:
\begin{equation}
M_X^w=\sum_{i\in J_X} |e_i\rangle\langle e_i|, 
\label{projection-manifest}
\end{equation}
where $J_X$ is the the set of indexes associated with the webpages that are manifest stories about $X$, where by ``manifest'' we mean stories that explicitly contain the word(s) ``X'' indicating the concept $X$, hence the superscript ``$w$'' in the notation, which stands for ``word.'' Indeed, we could as well have defined a more general projection operator $M_X^s=\sum_{i\in I_X} |e_i\rangle\langle e_i|$, onto the set of states that are stories about $X$ not necessarily of the manifest kind, {\it i.e.}, not necessarily containing the explicit word(s) indicating the concept(s) the stories are about, with $J_X\subset I_X$, and the superscript ``$s$'' now standing for ``story.'' 

To avoid possible confusions, we emphasize again the difference between the notion of \emph{state of a concept} and that of \emph{story about a concept}. The latter, in our definition, is a webpage, {\it i.e.}, a full-fledged printed or printable document. But webpages that are stories about a concept may explicitly contain the word indicating such concept, or not. For example, one can conceive a text explaining what \emph{Fruits} are, without ever writing the word ``fruits'' (using in replacement other terms, like ``foods in the same category of pineapple, pears and bananas''). On the other hand, the notion of state of a concept expresses a condition which cannot in general be reduced to that of a story, as it can also be a superposition of stories of that concept (or better, a superposition of the states associated with the stories of that concept), as expressed for instance in (\ref{general-state}) and (\ref{psiApsiB}), and a superposition of (states of) stories is not anymore a (state of a) story. 

Now, when considering a `tell a story measurement', we can also decide to only focus on stories having a predetermined content. In the double-slit experiment, this would correspond to only be interested in the detection of the electron by a certain subset of cells, indicated by a given set of indexes $J_X$, and not the others. More specifically, we can consider only those stories that are `stories about $X$', where $X$ is a given concept. This means that if the QWeb is in a pre-measurement state $|\psi_A\rangle$, which is the state of a given concept $A$, what we are asking through the measurement is if the stories about $X$ are good representatives of $A$ in state $|\psi_A\rangle$ (in the same way we can ask if a certain subset of traces of impact, say those of the central fringe, is a good example of electrons passing through the left slit; see the discussion of Sec.~\ref{interrogative}). In other words, we are asking how much $|\psi_A\rangle$ is meaning connected to concept $X$, when the latter is in one of the maximally concrete states defined by the webpages that are `stories of $X$', or even more specifically `manifest stories of $X$'. 

In the latter case, we can test this by using the projection operator $M_X^w$ and the Born rule. According to (\ref{Born}), the probability $\mu_A$ with which the concept $A$ in state $|\psi_A\rangle$ is evaluated to be well represented by a `manifest story about $X$', is given by the average:
\begin{equation}
\mu_A(i\in J_X)= \langle\psi_A| M_X^w |\psi_A\rangle=\sum_{i\in J_X} |\langle e_i|\psi_A\rangle|^2 = \sum_{i\in J_X}a_i^2, 
\label{muA}
\end{equation}
where for the last equality we have used (\ref{psiApsiB}). If we additionally assume that $A$ is more specifically described by a state that is a superposition only of those stories that explicitly contains the words ``A'' (manifest stories about $A$), the above probability becomes (omitting from now on the argument, to simplify the notation): $\mu_A= \sum_{i\in J_{A,X}}a_i^2$, where $J_{A,X}$ denotes the sets of indexes associated with the webpages jointly containing the words ``A'' and ``X.'' Note that if $n_{A,X}=|J_{A,X}|$ is the number webpages containing both terms ``A'' and ``X,'' $n_A=|J_A|$ and $n_X=|J_X|$ are the webpages containing the ``A'' term and the ``X'' term, respectively, we have $n_{A,X}\leq n_A$ and $n_{A,X}\leq n_X$. Becoming even more specific, we can consider states of $A$ expressing a uniform meaning connection towards all the different manifest stories about $A$, that is, characteristic function states of the form: 
\begin{equation}
|\chi_A\rangle={1\over \sqrt{n_A}} \sum_{j\in J_A} e^{i\alpha_j}|e_j\rangle,
\label{characteristicA}
\end{equation}
for which the probability (\ref{muA}) becomes: 
\begin{equation}
\mu_A= \langle\chi_A| M_X^w |\chi_A\rangle=\sum_{i\in J_{A,X}}{1\over n_A}= {n_{A,X}\over n_A},
\label{muA-bis}
\end{equation}
which can be simply interpreted as the probability of randomly selecting a webpage containing the term ``X,'' among those containing the terms ``A.'' 

With respect to the double-slit experiment analogy, the probability $\mu_A$ describes the ``only left slit open'' situation, and of course, \emph{mutatis mutandis}, we can write (with obvious notation) an equivalent expression for a different concept $B$: $\mu_B= \langle\chi_B| M_X^w |\chi_B\rangle=\sum_{i\in J_{B,X}}{1\over n_B}= {n_{B,X}\over n_B}$. So, when calculating the probability $\mu_{AB}$ for the combination $AB$ of the two concepts $A$ and $B$, we are in a situation equivalent to when the two slits are kept jointly open, with the question asked being now about the meaning connection between $AB$, in state $|\psi_{AB}\rangle$, and a (here manifest) story about $X$. Concerning the state $|\psi_{AB}\rangle$, describing the combination, we want it to be able to account for the emergence of meanings that can possibly arise when the two concepts $A$ and $B$ are considered one in the context of the other, and for consistency reasons we expect the probability $\mu_{AB}$ to be equal to ${n_{AB,X}\over n_{AB}}$ (since we are here limiting our discussion, for simplicity, to manifest stories), where $n_{AB}$ is the number of webpages containing both the ``A'' and ``B'' terms and $n_{AB,X}$ is the number of webpages containing in addition also the ``X'' term, and of course: 
$n_{AB,X}\leq n_{AB}$, $n_{AB}\leq n_{A}$ and $n_{AB}\leq n_{B}$. This can be easily achieved if the state of $AB$ is taken to be the characteristic function state: $|\chi_{AB}\rangle={1\over \sqrt{n_{AB}}} \sum_{j\in J_{AB}} e^{i\delta_j}|e_j\rangle$, however, coming back to our discussion of Sec.~\ref{double-slit}, this would not be a satisfactory way to proceed, as the modeling would then remain at the level of the canvas of printed documents of the Web, and would therefore not be able to capture the level of meaning associated with it, that is, the more abstract QWeb entity. It is only at the level of the latter that emergent meanings can be explained as the result of combining concepts.

By analogy with the paradigmatic double-slit experiment, we will here assume that a state of $AB$, {\it i.e.}, a state of the combination of the two concepts $A$ and $B$, when they are in individual states $|\psi_A\rangle$ and $|\psi_B\rangle$, respectively, can be generally represented as a superposition vector (\ref{superposition}). Since here we are considering the special case where these states are characteristic functions, we more specifically have: 
\begin{equation}
|\psi_{AB}\rangle= {1\over\sqrt{2}}(|\chi_A\rangle + |\chi_B\rangle),
\end{equation} 
where we have assumed for simplicity that $|\chi_A\rangle$ and $|\chi_B\rangle$ can be taken to be orthogonal states (this needs not to be the case  in general). The interference contribution ${\rm Int}_{AB}= \Re\, \langle \chi_A |M_X^w|\chi_B\rangle$ can then be calculated by observing that: 
\begin{eqnarray}
M^w_X |\chi_B\rangle&=&\left(\sum_{j\in J_X} |e_j\rangle\langle e_j| \right)\left({1\over \sqrt{n_B}} \sum_{k\in J_B} e^{i\beta_k}|e_k\rangle\right)\nonumber\\
&=& {1\over \sqrt{n_A}}\sum_{j\in J_X} \sum_{k\in J_B}e^{i\beta_k}|e_j\rangle\underbrace{\langle e_j|e_k\rangle}_{\delta_{jk}}={1\over \sqrt{n_B}}\sum_{j\in J_{B,X}}e^{i\beta_j}|e_j\rangle,
\label{Mwchi}
\end{eqnarray}
so that, multiplying the above expression from the left by $\langle \chi_A |$ and taking the real part, we obtain: 
\begin{eqnarray}
{\rm Int}_{AB}&=&\Re\left({1\over \sqrt{n_A}}\sum_{j\in J_A} e^{-i\alpha_j}\langle e_j|\right)\left({1\over \sqrt{n_B}}\sum_{k\in J_{B,X}}e^{i\beta_k}|e_k\rangle\right)\nonumber\\
&=& {1\over \sqrt{n_An_B}}\sum_{j\in J_A}\sum_{k\in J_{B,X}}\underbrace{\langle e_j|e_k\rangle}_{\delta_{jk}} \underbrace{\Re\, e^{i(\beta_k-\alpha_j)}}_{\cos(\beta_k-\alpha_j)}=\sum_{j\in J_{AB,X}} {\cos(\beta_j-\alpha_j)\over \sqrt{n_An_B}}.
\label{Interf}
\end{eqnarray}
According to (\ref{interference}), (\ref{muA-bis}) and (\ref{Interf}), the probability $\mu_{AB}$ for the combined concept $AB$ is therefore:
\begin{equation}
\mu_{AB}= {1\over 2}\Big(\underbrace{n_{A,X}\over n_A}_{\mu_A} +\underbrace{n_{B,X}\over n_B}_{\mu_B}\Big)+\sum_{j\in J_{AB,X}} {\cos(\beta_j-\alpha_j)\over \sqrt{n_An_B}}.
\label{muAB-nocontext}
\end{equation}

It is important to observe in (\ref{muAB-nocontext}) the role played by the phases $\alpha_j$ and $\beta_j$ characterizing the states $|\chi_A\rangle$ and $|\chi_B\rangle$. When they are varied, the individual probabilities $\mu_A$ and $\mu_B$ remain perfectly invariant, whereas the values of $\mu_{AB}$ can explore an entire range of values, within the interference interval $I_{AB}= [\mu^{\rm min}_{AB},\mu^{\rm max}_{AB}]$, where according to (\ref{muAB-nocontext}) we have: 
\begin{equation}
\mu^{\rm min}_{AB}= {1\over 2}\left({n_{A,X}\over n_A}+{n_{B,X}\over n_B}\right)-{n_{AB,X}\over \sqrt{n_An_B}},\quad\quad \mu^{\rm max}_{AB}= {1\over 2}\left({n_{A,X}\over n_A}+{n_{B,X}\over n_B}\right)+{n_{AB,X}\over \sqrt{n_An_B}}.
\label{muAB-interval}
\end{equation}
Therefore, we see that via the interference effects, the co-occurrence of the terms ``A,'' ``B'' and ``X'' is independent of what is revealed in the Web for the co-occurrence of just ``A'' and ``X,'' or the co-occurrence of just ``B'' and ``X''. This means that it is really at the more abstract level of the QWeb, and not of the Web, that these three situations of co-occurrence can be seen to be related to each other.

\section{Adding context}
\label{context}

According to (\ref{muAB-interval}), by using the superposition principle and the corresponding interference effects, we can extend the values of the probability $\mu_{AB}$ beyond those specified by the uniform average $\mu_{AB}^{\rm uni}={1\over 2}({n_{A,X}\over n_A}+{n_{B,X}\over n_B})$. One may wonder then if, generally speaking, interference effects would be sufficient to model all possible situations. The answer is negative, and to see why let us consider a simple  example of a collection of documents for which interference effects are insufficient for their modeling.\footnote{The example is taken from \cite{TCS2017}. Note however that the two situations described in \cite{TCS2017} required both the use of `interference plus context effects', contrary to what was stated in the article. Here we provide a corrected version of the example, where the first situation only requires interference effects, whereas the second situation requires interference plus context effects.}

\begin{figure}[htbp]
\begin{center}
\includegraphics[width=13cm]{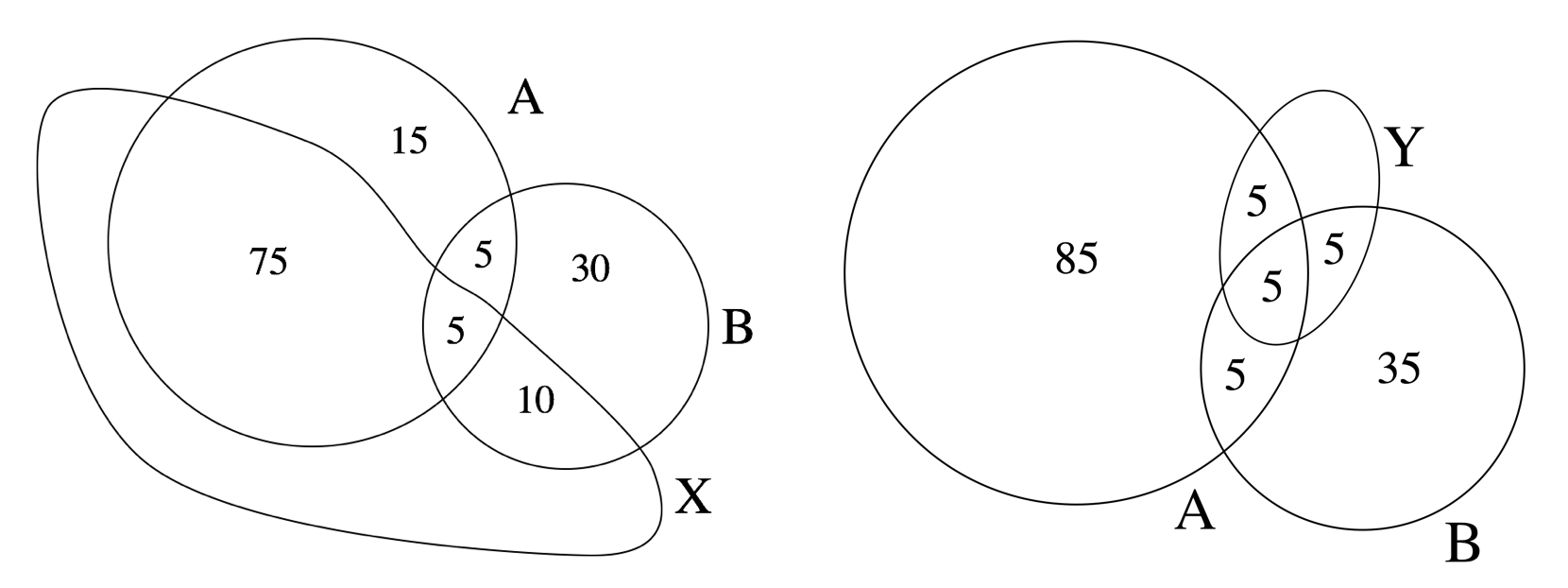}
\caption{A schematic Venn-diagram representation of the number of documents containing the words ``A,'' ``B'' and ``X'' (left) which can be modeled using only intereference effects, and the words ``A,'' ``B'' and  ``Y'' (right), which instead also require context effects.} 
\label{Figure5}
\end{center}
\end{figure}

Assume that the collection is formed by $n$ documents ($n\geq 140$), that $n_A=100$ of them contain a given word ``A,'' and $n_B=50$ of them contain another word ``B.'' Also, the number of documents containing both words is assumed to be $n_{AB}=10$ (See Fig.~\ref{Figure5}). Consider then a third word ``X,'' which is assumed to be present in $80$ of the documents containing the word ``A,'' in $15$ of the documents containing the word ``B,'' and in $5$ of the documents containing both words, that is: $n_{A,X}=80$, $n_{B,X}=15$, $n_{AB,X}=5$. So, $\mu_A={n_{A,X}\over n_A}={80\over 100}=0.8$, $\mu_B={n_{B,X}\over n_B}={15\over 50}=0.3$ and $\mu_{AB}^{\rm uni}={1.1\over 2}= 0.55$. We also have, ${n_{AB,X}\over \sqrt{n_An_B}}={5\over \sqrt{5000}}\approx 0.07$, so that  $\mu^{\rm min}_{AB}\approx 0.55-0.07=0.48$ and $\mu^{\rm max}_{AB}\approx 0.55+0.07=0.62$. 

Now, as we said, $\mu_{AB}$, for consistency reasons, should be equal to ${n_{AB,X}\over n_{AB}}={5\over 10}=0.5$, {\it i.e.}, to the probability of randomly selecting a document containing the word ``X,'' among those containing the words ``A'' and ``B.'' Since $0.5$ is contained in the interference interval $I_{AB}=[0.48,0.62]$, by a suitable choice the phase differences in (\ref{muAB-nocontext}), the equality $\mu_{AB} = {n_{AB,X}\over n_{AB}}$ can be obtained, hence interference effects are sufficient to model this situation.  But if we consider a word  ``Y'' that, different from ``X,'' would only be present in $10$ of the documents containing the word ``A'' and in $10$ of those containing the word ``B'' (See Fig.~\ref{Figure5}), this time we have $\mu_A={n_{A,Y}\over n_A}={10\over 100}=0.1$, $\mu_B={n_{B,Y}\over n_B}={10\over 50}=0.2$ and $\mu_{AB}^{\rm uni} ={0.3\over 2}= 0.15$. So, $\mu^{\rm min}\approx 0.15-0.07=0.08$ and $\mu^{\rm max}\approx 0.15+0.07=0.22$, which means that ${n_{AB,Y}\over n_{AB}}=0.5$ is not anymore contained in the interference interval $I_{AB}=[0.08,0.22]$. Hence, interference effects are not sufficient to model this situation. 

Additional mechanisms should therefore be envisioned to account for all the probabilities that can be calculated by counting the relative number of documents containing certain words and co-occurrences of words. A possibility is to explore more general forms of measurements on more general versions of the QWeb entity. In our approach here, we focused on the superposition principle to account for the emergence of new meanings when concepts are combined. But of course, when a cognitive entity interacts with a meaning entity, the emergence of meaning is not the only element that might play a role. In human reasoning for instance, a two-layer structure can be evidenced: one consisting of \emph{conceptual thoughts}, where a combination of concepts is evaluated as a new single concept, and the other consisting of \emph{classical logical thoughts}, where a combination of concepts is evaluated as a classical combinations of different entities \cite{asv2015}. 

To also account for the existence of classical logical reasoning, one can define more general `tell a story measurements', by considering a specific type of Hilbert space called \emph{Fock space}, originally used in \emph{quantum field theory} to describe situations where there is a variable number of identical entities. This amounts considering the QWeb as a more general ``quantum field entity'' that can be in different \emph{number operator} states and in different superpositions of these states. In the present case, since we are only considering the combination of two concepts, the construction of the Fock space ${\cal F}$ can be limited to two sectors: ${\cal F}= {\cal H}\oplus ({\cal H}\otimes {\cal H})$, where ``$\oplus$'' denotes a \emph{direct sum} between the first sector ${\cal H}$ (isomorphic to $\compl^n$) and the second sector ${\cal H}\otimes {\cal H}$ (isomorphic to $\compl^{2n}$), where ``$\otimes$'' denotes the \emph{tensor product}. The first sector describes the one-entity states, where the combination of the two concepts $A$ and $B$ is evaluated as a new (emergent) concept, typically described by a superposition state (\ref{superposition}). The second sector describes the two-entity situation, where the two concepts $A$ and $B$ remain separate in their combination, which is something that can be described by a so-called \emph{product} (non-entangled) state $|\psi_A\rangle\otimes |\psi_B\rangle$. 

Instead of (\ref{superposition}), we can then consider the more general superposition state:
\begin{equation}
|\psi_{AB}\rangle= \sqrt{1-m^2}\, e^{i\nu}{1\over\sqrt{2}}(|\psi_A\rangle + |\psi_B\rangle) + m\, e^{i\lambda} |\psi_A\rangle\otimes |\psi_B\rangle,
\label{superposition-fock}
\end{equation}
where the number $0\leq m\leq 1$ determines the degree of participation in the second sector. Also, instead of (\ref{projection-manifest}), we have to consider a more general projection operator, acting now on both sectors. Here we can distinguish the two paradigmatic projection operators: 
\begin{equation}
M^{w,{\rm and}}_X= M_X^w\oplus(M_X^w \otimes M_X^w), \quad M^{w,{\rm or}}_X= M_X^w\oplus(M_X^w \otimes \mathbb{I} + \mathbb{I} \otimes M_X^w -M_X^w \otimes M_X^w),
\label{and-or-projections}
\end{equation}
where $M^{w,{\rm and}}_X$ describes the situation where the combination of concepts is logically evaluated as a \emph{conjunction} (and), whereas $M^{w,{\rm or}}_X$ describes the situation where the combination of concepts is logically evaluated as a \emph{disjunction} (or). When we use $M^{w,{\rm and}}_X$, one finds, in replacement of (\ref{interference}), the more general formula:\footnote{It is not in the scope of the present chapter to enter into the details of this Fock space modeling and we simply refer the interested reader to \cite{a2009a,asv2015b,as2017}.}
\begin{equation}
\mu_{AB}=m^2\, \mu^{\rm and}_{AB} +(1-m^2)\left[{1\over 2}(\mu_A + \mu_B)+{\rm Int}_{AB}\right],
\label{interference-fock}
\end{equation}
where $\mu^{\rm and}_{AB} = \mu_A\mu_B$. However, this will not be sufficient to model all possibe data, as is clear that in the previously mentioned example of word ``Y,'' we have: $\mu^{\rm and}_{AB}=0.02$, so that the interval of values that can be explored by the above convex combination (by varying not only the phases $\alpha_j$ and $\beta_j$, but now also the coefficient $m$) is $[0.02, 0.22]$, which still doesn't contain the value $0.5$ of ${n_{AB,Y}\over n_{AB}}$. When we use instead $M^{w,{\rm or}}_X$, we have to replace $\mu^{\rm and}_{AB}$ in (\ref{interference-fock}) by $\mu^{\rm or}_{AB}=\mu_A +\mu_B - \mu_A\mu_B$, whose value for the word ``Y'' of our example is $0.28$, so that the interval of possible values becomes $[0.08, 0.28]$, which however is still not sufficient.

\begin{figure}
\begin{center}
\includegraphics[width=6cm]{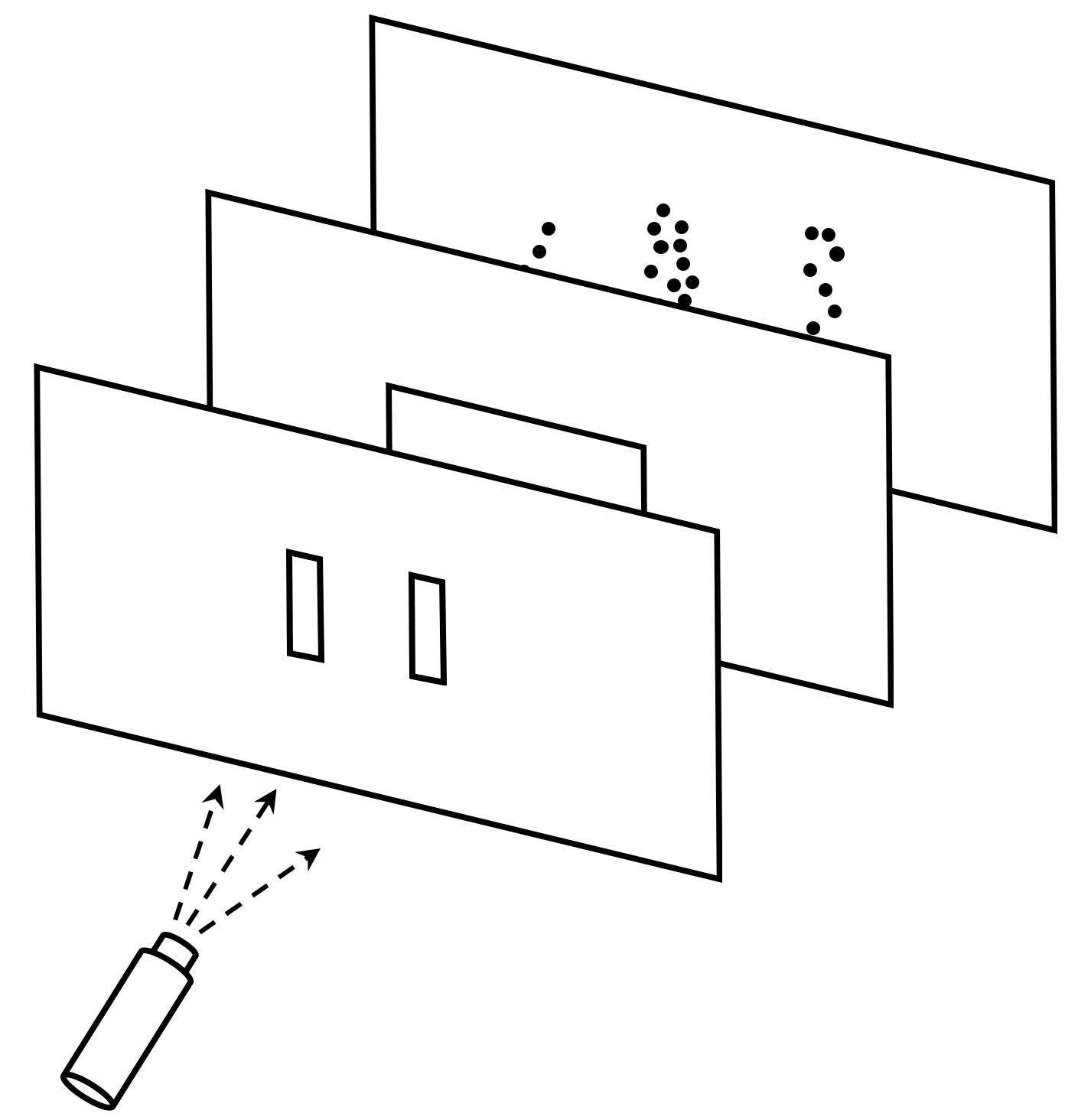}
\caption{By placing a screen with a mask (and more generally a filter) between the barrier and the detection screen, the structure of the observed interference pattern can be modulated. The effect of this additional structure can be ideally described using a projection operator.}
\label{Figure4}
\end{center}
\end{figure}

So, we must find some other cognitive effects, in order to be able to model and provide an explanation for a wide spectrum of experimental values for the probabilities, related to different possible collections of documental entities. A general way of proceeding, remaining in a ``first sector'' modeling of the QWeb, is to consider that there would be also \emph{context effects} that can alter the QWeb state before it is measured. In the double slit experiment analogy, we can imagine a mask placed somewhere in between the barrier and the screen, acting as a filter allowing certain states to pass through whereas others will be  blocked (see Fig.~\ref{Figure4}).
Note that if we place the mask close to the detection screen, some cells will be deactivated, as the components of the pre-measurement state relative to them will  be filtered out by the mask. On the other hand, if it is placed close to the double-slit barrier, it will allow to control the transmission through the slits and produce, by changing its position, a continuum of interference figures, for instance interpolating the probability distributions of the two one-slit arrangements; see \cite{Bachetal2013}. More complex effects can of course be obtained if the mask is placed at some finite distances from the barrier and screen, and more general filters than just masks can also be considered, but their overall effect will always be that only certain states will be allowed to interact with the measuring apparatus (here the screen). 

From a cognitivistic viewpoint, context effects can have different origins and logics. For instance, we can consider that an interrogative context, for the very fact that a given question is asked, will inevitably alter the state of the meaning entity under consideration. Even more specifically, consider the example of a cognitive entity that is asked to tell a story (it can be a person, a search engine, or the combination of both). For this, a portion of the entity's memory needs to become accessible, and one can imagine that the extent and nature of such available portion of memory can depend on the story that is being asked.\footnote{In the IR ambit, this can also be associated with constraints related to geographical locations and search histories \cite{Melucci2012, melucci2015}.}

So, we will now assume that when the QWeb entity is subjected to a `tell a story measurement', there will be a preliminary change of state, and we will adopt the very simple modeling of such state change by means of an orthogonal projection operator, which in general can also depend on the choice of stories we are interested in, like `stories about $X$', so we will generally write $N_X$ for it ($N_X^2=N_X=N_X^\dagger$). Just to give a simple example of a $X$-dependent projection $N_X$, it could be taken to be the projection operator onto the subspace of QWeb's states that are `states of $X$' (we recall that a `state of $X$' is generally not necessarily also a `story about $X$'). However, in the following we will just limit ourselves to the idealization that context effects can be formally modeled using a projection operator, without specifying their exact nature and origin. So, the presence of this additional context produces the pre-measurement transitions: $|\psi_A\rangle\to |\psi'_A\rangle$, $|\psi_B\rangle\to |\psi'_B\rangle$ and $|\psi_{AB}\rangle\to |\psi'_{AB}\rangle$, where we have defined (from now on, for simplicity, we just write $N$ for $N_X$, dropping the $X$-subscript):
\begin{equation}
|\psi'_A\rangle = {N|\psi_A\rangle \over \|N|\psi_A\rangle\|}, \quad |\psi'_B\rangle = {N|\psi_B\rangle \over \|N|\psi_B\rangle\|},\quad |\psi'_{AB}\rangle = {N|\psi_{AB}\rangle \over \|N|\psi_{AB}\rangle\|}.
\end{equation}
With the above re-contextualized states, the probability $\mu_A = \langle\psi'_A| M_X^w |\psi'_A\rangle$  becomes: 
\begin{equation}
\mu_A = {\langle\psi_A | N^\dagger M_X^w N|\psi_A\rangle\over \|N|\psi_A\rangle\|^2}={\langle\psi_A | N M_X^w N|\psi_A\rangle\over \langle\psi_A | N |\psi_A\rangle}={1\over p_A}\langle\psi_A | N M^w N|\psi_A\rangle,
\label{muAandmuB}
\end{equation}
where for the second equality we have used $\|N|\psi_A\rangle\|^2=\langle \psi_A |N^\dagger N|\psi_A\rangle=\langle \psi_A |N^2|\psi_A\rangle=\langle\psi_A | N |\psi_A\rangle$, and for the last equality we have defined the probability $p_A=\langle\psi_A | N |\psi_A\rangle$, for the state $|\psi_A\rangle$ to be an eigenstate of the context $N$. Similar expressions clearly hold also for the concept $B$: $\mu_B = {1\over p_B}\langle\psi_B | N M^w N|\psi_B\rangle$, with $p_B=\langle\psi_B | N |\psi_B\rangle$, and for the probability $\mu_{AB}=\langle\psi'_{AB}| M_X^w |\psi'_{AB}\rangle$, relative to the concept combination $AB$, we now have: 
\begin{eqnarray}
\mu_{AB} &=& {\langle\psi_{AB} | N^\dagger M_X^w N|\psi_{AB}\rangle\over \|N|\psi_{AB}\rangle\|^2} = {\langle\psi_{AB} | NM_X^w N|\psi_{AB}\rangle\over \langle\psi_{AB} | N |\psi_{AB}\rangle}\nonumber \\
&=&{\langle\psi_A | NM_X^w N|\psi_A\rangle +\langle\psi_B | NM_X^w N|\psi_B\rangle + 2 \Re\, \langle\psi_A | NM_X^w N|\psi_B\rangle \over \langle\psi_A | N |\psi_A\rangle +\langle\psi_B | N |\psi_B\rangle + 2 \Re\, \langle\psi_A | N |\psi_B\rangle}\nonumber\\
&=&{p_A\,\mu_A +p_B\, \mu_B + 2 \Re\, \langle\psi_A | NM_X^w N|\psi_B\rangle \over p_A +p_B + 2 \Re\, \langle\psi_A | N |\psi_B\rangle}.
\label{muAB}
\end{eqnarray}

The first two terms at the numerator of (\ref{muAB}) correspond to a \emph{weighted average}, whereas the third term, both at the numerator and denominator, is the interference-like contribution. Note that in the special case where $|\psi_A\rangle$ and $|\psi_B\rangle$ are eigenstates of the context $N$, that is, $N|\psi_A\rangle=|\psi_A\rangle$ and $N|\psi_B\rangle=|\psi_B\rangle$, we have $p_A =p_B=1$, so that (\ref{muAB}) reduces to (\ref{interference}), or, if $|\psi_A\rangle$ and $|\psi_B\rangle$ are not orthogonal vectors, to: 
\begin{equation}
\mu_{AB}= {{1\over 2}(\mu_A +\mu_B) + \Re\, \langle\psi_A | M_X^w |\psi_B\rangle \over 1 + \Re\, \langle\psi_A |\psi_B\rangle},
\label{muAB-nocontext2}
\end{equation}
where the weighted average now becomes a uniform one. The more general expression (\ref{muAB}), incorporating both context and interference effects, allows to cover a much larger range of values. In fact, as we show in Appendix~\ref{context and interference}, under certain assumptions the full $[0,1]$ interval of values can be spanned, thus allowing all possible data about occurrence and co-occurrence of words to be modeled.

\section{Conclusion}\label{Conclusion}

In this chapter, we have motivated a fundamental distinction between the Web of printed pages (or any other collection of documental entities), and a more abstract entity of meaning associated with it, which we have called the QWeb, for which we have proposed a Hilbertian (Born rule based) quantum model. In our discussion, we have focused on an important class of measurements, which we have called the `tell a story measurements', whose outcome states are associated with the $n$ webpages and were taken to form a basis of the ($n$-dimensional) Hilbert space. We have tested the model by considering the specific situation where only stories manifestly containing the words denoting certain concepts are considered, in order to allow to relate the theoretical probabilities with those obtained by calculating the relative frequency of occurrence and co-occurrence of these words, which in turn depend on how much the associated concepts are meaning-connected. We have done so by also considering context effects, in addition to interference effects, the former being modeled by means of orthogonal projection operators and the latter by means of superposition states. Also, we have extensively used the double-slit experiment as a guideline to motivate the transmigration of fundamental notions from physics to human cognition and theoretical computer science. 

Note that more general models than those explored here can also be considered, exploiting more general versions of the quantum formalism, like the GTR-model and the extended Bloch representation of quantum mechanics \cite{asdbs2015,asdb2015b, AertsSassolideBianchi2014,AertsSassoli2016,asdbianchi2017}. Hence, the ``Q'' in ``QWeb'' refers to a \emph{quantum structure} that needs not to be understood in the limited sense of the standard quantum formalism. We have also mentioned in Sec.~\ref{context} the possibility of working in a multi-sector Fock space, as a way to extend the range of probabilities that can be modeled. However, we observed that not all values can be modeled in this way. Another direction that can be explored (as an alternative to context effects) is to consider states whose meaning connections are not necessarily uniform, although still localized within the sets $J_A$ and $J_B$. A further other direction is to consider \emph{step function} states extending beyond the manifest word subspaces. For example, states of the form: $|\psi_A^{\, a}\rangle= a\, |\chi_A\rangle + \bar{a}\, |\bar{\chi}_A\rangle$, where $|\bar{\chi}_A\rangle={1\over\sqrt{n-n_A}} \sum_{j\notin J_A}e^{i\alpha_j}|e_j\rangle$, and $|a|^2 + |\bar{a}|^2 =1$.

Regarding the co-occurrences of words in documents, it is worth observing that they are determined by the meaning carried by the corresponding concepts and documents, and not by the physical properties of the latter. This means that we can access the traces left by meaning by analyzing the co-occurrence of words in the different physical (printed, or stored in memory) documents, and that such meaning ``stick out'' from the latter in ways that can be accessed without the intervention of the human minds that created it. Note however that the meaning extending out of these documents, here the webpages, is not the full meaning of the 
QWeb, as encoded in its quantum state. This is so because one cannot reconstruct the pre-measurement state of a quantum measurement by having only access to the outcome (collapsed) states and the associated probabilities of a single measurement. For this, one needs to perform a series of different measurements, characterized by different \emph{informationally complete} bases, as is done in so-called \emph{quantum state tomography} \cite{dariano2004}. Here we only considered the basis associated with the webpages, and it is still unclear which complementary measurements could be defined, using different bases and having a clear operational meaning, that is, which can be concretely performed, at least in principle \cite{TCS2017}.

Let us also observe that, generally speaking, in IR situations also the modeling of how human minds interact with the QWeb can and will play a role, in addition to the modeling per se of the QWeb. Indeed, as we mentioned already in Sec.~\ref{interrogative}, the outcome provided by a measurement of the QWeb, say a given story in a `tell a story measurement', becomes the input with which human minds will have to further interact with, which again can be described as a deterministic or indeterministic context, possibly creating new meanings. The formalism of quantum theory can again be used to model these human cognitive interactions, which is what is typically investigated in cognitive psychology experiments and again modeled using the mathematical formalism of quantum theory, in the emerging field known as \emph{quantum cognition}; see \cite{assv2018} and the references cited therein. 

We stress that, in our view, it is only when a more abstract -- meaning oriented -- approach is adopted in relation to documental entities, like the Web, and  an operational-realistic modeling of its conceptual structure is attempted, exploiting the panoply of quantum effects that have been discovered in the physics' laboratories, that, quoting from \cite{TCS2017}: ``a deeper understanding of how meaning can leave its traces in documents can be accessed, possibly leading to the development of more context-sensitive and semantic-oriented information retrieval models.'' Note however that we have not attempted here any evaluation of what are the pros and cons, differences and similarities, of our modeling and the other existing approaches, also integrating quantum features. Let us just mention, to give a few examples, Foskett's work in the eighties of last century \cite{Foskettetal1980}, Agosti \emph{et al.} work in the nineties  \cite{Agostietal1991,Agostietal1995}, and Sordoni \emph{et al.} more recent work, where the double-slit experiment analogy is also used to investigate quantum interference effects for topic models such as LDA \cite{Sordoni2013}.\footnote{Sordoni \emph{et al.} represented documents as superposition of topics, whereas in our approach documents are considered to be outcomes of the `tell a story' measurements. In other words, for Sordoni \emph{et al.}  a document is like an electron entering the double-slit apparatus, and the terms like the traces of impact on the detection screen. This is different from our perspective, where documents are instead the traces of impact on the detection screen and the equivalent of the electron entity is the QWeb entity.} 

To conclude, let us observe that in the same way the quantum cognition program, and its effectiveness, does not require the existence of microscopic quantum processes in the human brain \cite{assv2018}, the path ``towards a quantum Web'' that we  have sketched here, and in \cite{TCS2017}, where the Web of written documents is viewed as a ``collection of traces'' left by an abstract meaning entity -- the QWeb,  -- should not be confused with the  path ``towards a quantum Internet" \cite{duretal2017}, which is about constructing an Internet able to transmit ``quantum information,'' instead of just ``classical information,'' that is, information carried by entities allowing  quantum superposition to also take place and be fully exploited. In the future, there will certainly be a Quantum Internet and a Quantum Web, that is, there will be a physical Internet more and more similar in structure to the abstract Web of meanings it conveys. These will be fascinating times for the evolution of the human race on this planet, who will then be immersed in a fully developed \emph{noosphere}, but at the moment we are not there yet.

\appendix
\section{Interference plus context effects}\label{context and interference}

In this appendix, we show that using the ``interference plus context effects'' formula (\ref{muAB}), all data can in principle be modeled, by suitably choosing the different parameters. For simplicity, we start by assuming that $M^w_XN=NM^w_X$, {\it i.e.}, that $N$ and $M^w_X$ are compatible, so that the projection $N^\dagger M_X^w N$ can be simply written as $NM_X^w$, as is clear that $N^\dagger M_X^w N=N^\dagger N M_X^w=N^2M_X^w= NM_X^w$. In other words, we have $(NM^w)^\dagger=
NM^w$ and $(NM^w)^2=NM^w$. This means that we can define the following three orthogonal projectors: 
\begin{equation}
P_1=M_X^wN,\quad \quad P_2=(\mathbb{I} -M_X^w)N,\quad \quad P_3=\mathbb{I}-N,
\end{equation}
which are orthogonal to each other:
\begin{eqnarray}
&&P_1P_2 = M_X^wN(\mathbb{I} -M_X^w)N = M_X^wN^2 - (M_X^wN)^2 = 0,\nonumber\\
&&P_1P_3 = M_X^wN(\mathbb{I}-N)=M_X^wN-M_X^wN^2=0,\nonumber\\
&&P_2P_3 =(\mathbb{I} -M_X^w)N(\mathbb{I}-N)=(\mathbb{I} -M_X^w)(N-N^2)=0.
\end{eqnarray}
Consequently, we can write the Hilbert space as the direct sum: ${\cal H}={\cal H}_1 \oplus {\cal H}_2\oplus {\cal H}_3$, where ${\cal H}_1= P_1 {\cal H}$, ${\cal H}_2= P_2 {\cal H}$ and ${\cal H}_3= P_3 {\cal H}$ are three orthogonal subspaces, and we can write $|\psi_A\rangle$ and $|\psi_B\rangle$ as linear combinations of vectors belonging to them:
\begin{eqnarray}
|\psi_A\rangle&=&ae^{i\alpha}|e\rangle+a'e^{i\alpha'}|e'\rangle+ a'' e^{i\alpha''} |e''\rangle, \nonumber\\
|\psi_B\rangle&=&be^{i\beta}|f\rangle+b'e^{i\beta'}|f'\rangle+b'' e^{i\beta''} |f''\rangle, 
\label{expansions}
\end{eqnarray}
where $|e\rangle,|f\rangle$ are unit vectors in ${\cal H}_1$, $|e'\rangle,|f'\rangle$ are unit vectors in ${\cal H}_2$, and $|e''\rangle,|f''\rangle$ are unit vectors in ${\cal H}_3$. 
Considering that the vectors in the expansions (\ref{expansions}) are mutually orthogonal, it follows that:
\begin{eqnarray}
&&p_A\mu_A=\langle\psi_A | NM_X^w N |\psi_A\rangle = \langle\psi_A | P_1 |\psi_A\rangle =\langle\psi_A | ( ae^{i\alpha}|e\rangle ) =a^2,\nonumber\\
&& p_B\mu_B=\langle\psi_B | NM_X^w N |\psi_B\rangle = \langle\psi_B | P_1 |\psi_B\rangle =\langle\psi_B | ( be^{i\beta}|f\rangle ) =b^2.
\end{eqnarray}
We also have: 
\begin{eqnarray}
\Re\, \langle\psi_A | NM^wN |\psi_B\rangle &=& \Re\, \langle\psi_A | P_1 |\psi_B\rangle = \Re\, (\langle\psi_A | P_1)(P_1 |\psi_B\rangle) = \Re\, (\langle e|ae^{-i\alpha})(be^{i\beta}|f\rangle)\nonumber\\
&=& ab\, \Re\, e^{i(\beta-\alpha)} \langle e|f\rangle = abc\,\Re\, e^{i(\gamma+\beta-\alpha)}= abc \cos\phi,
\end{eqnarray}
where for the second equality we have used $P_1=P_1^2$, and for the fifth equality we have defined the positive number $c$ and the phase $\gamma$ such that $c\, e^{i\gamma} = \langle e|f\rangle$, whereas for the last equality we have defined $\phi =\gamma +\beta-\alpha$. In a similar way, we set $c'e^{i\gamma'} = \langle e'|f'\rangle$ and $\phi' =\gamma'+ \beta'-\alpha$, and considering that $N=\mathbb{I} N=[M_X^w + (\mathbb{I}- M_X^w)]N= P_1 +P_2$, we have:
\begin{equation}
\Re\, \langle\psi_A | N |\psi_B\rangle = \Re\, \langle\psi_A | P_1 |\psi_B\rangle +\Re\, \langle\psi_A |P_2 |\psi_B\rangle = abc \cos\phi+a'b'c'\cos\phi'.
\end{equation}
In a similar way, we have: 
\begin{eqnarray}
&&p_A = \langle\psi_A | N |\psi_A\rangle = \langle\psi_A | P_1 |\psi_A\rangle + \langle\psi_A |P_2 |\psi_A\rangle = a^2 + {a'}^2\nonumber\\
&&p_B = \langle\psi_B | N |\psi_B\rangle = \langle\psi_B | P_1 |\psi_B\rangle + \langle\psi_B |P_2 |\psi_B\rangle = b^2 + {b'}^2,
\end{eqnarray}
from which it follows that:
\begin{equation}
{a'}^2 = p_A-a^2 = p_A(1-\mu_A)=p_A\bar{\mu}_A,\quad\quad {b'}^2 = p_B-b^2 = p_B(1-\mu_B) = p_B\bar{\mu}_B,
\end{equation}
where we have defined $\bar{\mu}_A =1-\mu_A$ and $\bar{\mu}_B =1-\mu_B$. We can thus rewrite (\ref{muAB}) as:
\begin{equation}
\mu_{AB} ={p_A\,\mu_A +p_B\,\mu_B + 2\sqrt{p_Ap_B}\sqrt{\mu_A\mu_B} \, c \cos\phi \over p_A+p_B+ 2\sqrt{p_Ap_B}\, (\sqrt{\mu_A\mu_B} \, c \cos\phi+\sqrt{\bar{\mu}_A\bar{\mu}_B} \, c' \cos\phi')}.
\label{muAB-bis}
\end{equation}

To relate (\ref{muAB-bis}) to the webpages' counts, we consider the situation where states are uniform superpositions of states associated with manifest stories (characteristic function states). Different from the ``only interference effects situation'' of Sec.~\ref{modelWeb}, we however now assume that the vectors represented by characteristic functions are those that are obtained following the action of the context $N$. Clearly, this should only be considered as a rough approximation meant to illustrate that the present approach can handle the probabilities calculated by performing webpages' counts.
So, we assume that $|\psi'_A\rangle = |\chi_A\rangle$ and $|\psi'_B\rangle = |\chi_B\rangle $, so that according to (\ref{muA-bis}), (\ref{muAB-bis}) can be written: 
\begin{equation}
\mu_{AB} ={p_A\,{n_{A,X}\over n_A} +p_B\,{n_{B,X}\over n_B} + 2\sqrt{p_Ap_B}\sqrt{{n_{A,X}n_{B,X}\over n_An_B}} \,c \cos\phi \over p_A+p_B+ 2\sqrt{p_Ap_B}\left(\sqrt{{n_{A,X}n_{B,X}\over n_An_B}} \,c \cos\phi+\sqrt{{n_{A,X'}n_{B,X'}\over n_An_B}} \,c' \cos\phi'\right)},
\label{muAB-tris}
\end{equation}
where we have defined $n_{A,X'}= n_A - n_{A,X}$ and $n_{B,X'}=n_B - n_{B,X}$, which are the number of webpages containing the term ``A'' but not the term ``X,'' and the term ``B'' but not the term ``X,'' respectively. The consistency of the model is therefore about finding values for $p_A, p_B, c, c' \in [0,1]$ and $\phi, \phi' \in [0, 2\pi]$, such that (\ref{muAB-tris}) can be equal to ${n_{AB,X}\over n_{AB}}$. This will always be the case since (\ref{muAB-bis}) can in fact deliver all values between $0$ and $1$, as we are now going to show. 

Consider first the limit case where (\ref{muAB-bis}) is equal to $0$. Then its numerator has to vanish. If, say, we choose $c=1$ and $\phi=\pi$, this means that we must have $(\sqrt{p_A\,\mu_A}-\sqrt{p_B\,\mu_B})^2=0$, which is satisfied if ${p_A\over p_B}={\mu_B\over \mu_A}$. For the other limit case where (\ref{muAB-bis}) is equal to $1$, if we choose $c'=1$ and $\phi'=\pi$, we have the condition: $(\sqrt{p_A\,\bar{\mu}_A}-\sqrt{p_B\,\bar{\mu}_B})^2=0$, which is clearly satisfied if ${p_A\over p_B}={\bar{\mu}_B\over \bar{\mu}_A}$. For the intermediate values between $0$ and $1$, if we set $\phi=\phi'={\pi \over 2}$ (no-interference condition), (\ref{muAB-bis}) becomes: 
\begin{equation}
\mu_{AB} ={p_A\over p_A+p_B}\,\mu_A +{p_B\over p_A+p_B}\,\mu_B,
\label{muAB-4}
\end{equation}
which is a convex combination of $\mu_A$ and $\mu_B$. Therefore, by varying $p_A$ and $p_B$, by just considering context effects all values contained in the interval $[\min(\mu_A,\mu_B), \max(\mu_A,\mu_B)]$ can be obtained. 

To be able to extend further the interval, the relative phases $\phi$ or $\phi'$ have to be allowed to take values different from ${\pi \over 2}$. In this way, also the intervals $[0,\min(\mu(A),\mu(B))]$ and $[\max(\mu(A),\mu(B)),1]$ can be reached. To see this, we have to study the behavior of $\mu_{AB}=\mu_{AB}(x,x')$ as a function of the two variables $(x,x')=(\cos\phi, \cos\phi')$. We know that $\mu(AB;0,0)$ is given by (\ref{muAB-4}), so we just have to show that, for suitable choices of $p_A$ and $p_B$, by varying $x$ and $x'$ we can reach the $0$ value. For a given $x$, $\mu_{AB}(x,x')$ monotonically decreases as $x'$ increases. Thus, we only have to consider $\mu_{AB}(x,1)$, and by studying the sign of $\partial_x\mu_{AB}(x,1)$  one can easily check that (we leave this as an exercise) $\mu_{AB}(x,1)$ monotonically increases with $x$. Thus, the minimum corresponds to $\mu_{AB}(-1,1)$, which is $0$ if $c=1$ and ${p_A\over p_B}={\mu_B\over \mu_A}$. Similarly, we can consider $\mu_{AB}(x,-1)$ and check that $\mu_{AB}(x,-1)$ also monotonically increases with $x$. Thus, its maximum corresponds to $\mu_{AB}(1,-1)$, which is $1$ if $c'=1$ and ${p_A\over p_B}={\bar{\mu}_B\over \bar{\mu}_A}$. In other words, for arbitrary $\mu_A$, $\mu_B$ and $\mu_{AB}$, a quantum representation that can faithfully model the experimental data exist, if both interference and context effects are considered.

\section{Meaning bond}\label{bond}
 
In this appendix, we offer a more specific interpretation for the normalized weights $a_j$ characterizing the linear combination in (\ref{psiApsiB}), in terms of a notion of \emph{meaning bond} of a concept with respect to another concept, when the QWeb is in a given state $|\psi\rangle$. For this, let $M_A$ and $M_B$ be the projection operators onto the set of QWeb states that are `states of $A$' and `states of $B$', respectively. We can then define the $\psi$-meaning bond $M_{\psi}(B|A)$ of $B$ towards $A$ by the ratio:
\begin{equation}
M_{\psi}(B|A)={p_{\psi}(B|A)\over p_{\psi}(B)},
\label{bond1}
\end{equation}
where $p_{\psi}(B)=\langle \psi|M_B |\psi \rangle$ is the probability for the QWeb's state $|\psi\rangle$ to be successfully tested as being also a `state of $B$', and 
\begin{equation}
p_{\psi}(B|A)={\langle \psi|M_AM_BM_A |\psi \rangle\over \langle \psi|M_A|\psi \rangle}
\label{conditional}
\end{equation}
is the conditional probability of having the QWeb's state being successfully tested as being a `state of $B$', when it has been successfully tested to be a `state of $A$'. Indeed, if the QWeb state $|\psi\rangle$ was successfully tested to be a `state of $A$', according to the \emph{projection postulate} the state immediately following the test is $|\psi_A\rangle ={M_A |\psi\rangle\over \| M_A|\psi\rangle \|}$, which is now a `state of $A$'. And we have $p_{\psi}(B|A)=\langle\psi_A| M_B|\psi_A\rangle$, hence (\ref{conditional}) possesses a sound interpretation as a conditional probability.

The $\psi$-meaning bond $M_{\psi}(A|B)$ of $A$ towards $B$ can be similarly obtained by interchanging in (\ref{bond1}) the roles of $A$ and $B$, and since in general $[M_A,M_B]\neq 0$, $M_{\psi}(A|B)\neq M_{\psi}(B|A)$, which means that the meaning bond of $A$ towards $B$ will not in general coincide with the meaning bond of $B$ towards $A$. So, if $p_{\psi_A}(B)$ and $p_{\psi}(B)$ are interpreted as measuring how much of the meaning of $B$ is present in the QWeb, when the latter is in state $|\psi_A\rangle$ and $|\psi \rangle$, respectively, it is clear that the meaning bond $M_{\psi}(B|A)={p_{\psi_A}(B)\over p_{\psi}(B)}$, being their ratio, it measures the relative increase or decrease of the meaning presence of $B$ when the QWeb state $|\psi \rangle$ is further contextualized by a concept $A$. In that respect, we can also say that if $B$ is more (resp., less) meaning present in the QWeb, when its state is further contextualized by a concept $A$, then for such state there is an attractive (resp., repulsive) meaning bond of $B$ towards $A$, whereas if $p_{\psi_A}(B)=p_{\psi}(B)$ the meaning bond can be said to be neutral. Also, since we have $p_{\psi_B}(B)=1$, the meaning bond of $B$ towards itself is $M_{\psi}(B|B)=p^{-1}_{\psi}(B)$, so that there will be self-neutrality when $p_{\psi}(B)=1$, and self-attraction if $p_{\psi}(B)<1$ (but there cannot be self-repulsion). 

We now observe that: $p_{\psi}(W_j)M_{\psi}(W_j|A)=p_{\psi}(W_j|A)=\langle\psi_A| P_j|\psi_A\rangle =a_j^2$, where $P_j=|e_j\rangle\langle e_j|$ is the projection operator onto the one-dimensional subspace generated by the `ground state of $W_j$', {\it i.e.}, of the story-concept indicated by the specific combination of words contained in the webpage ${\rm W}_j$. Thus, we have that the coefficients $a_j$ in the expansion of the state $|\psi_A\rangle={M_A|\psi\rangle\over \|M_A|\psi\rangle \|} = \sum_{j=1}^n a_je^{i\alpha_j}|e_j\rangle$, which is a `state of $A$', can be written:
\begin{equation}
a_j=\sqrt{p_{\psi}(W_j)M_{\psi}(W_j|A)}
\label{aj}
\end{equation}
and therefore are given by (the square root of) the `$\psi$-meaning bond of $W_j$ towards $A$', normalized by the probability $p_{\psi}(W_j)$, and in that sense we can say that they express a meaning connection between $A$ and the $W_j$. Note also that in the case where $|\psi\rangle$ corresponds to the uniform state $|\chi\rangle={1\over \sqrt{n}} \sum_{j=1}^n e^{i\rho_j}|e_j\rangle$, (\ref{bond1}) reduces to the ratio
\begin{equation}
M_{\chi}(B|A)={n\, n_{AB} \over n_An_B},
\label{bond2}
\end{equation}
which corresponds to the more specific notion of meaning bond introduced in \cite{a2011} (see also \cite{aabssv2016}).

\bigskip
\noindent
{\bf Acknowledgments.} This work was supported by the Marie Sk{\l}odowska-Curie Innovative Training Network 721321 -- ``QUARTZ''.


\begin{thebibliography}{99}

\bibitem{rijsbergen2004} Van Rijsbergen, C. J. (2004). {\it The Geometry of Information Retrieval}. Cambridge: Cambridge University Press.

\bibitem{asdbs2015} Aerts, D., Sassoli de Bianchi, M. \& Sozzo, S. (2016). On the foundations of the Brussels operational-realistic approach to cognition,  {\it Frontiers in Physics 4}, 17, doi: 10.3389/fphy.2016.00017.

\bibitem{aerts-etal2018} Aerts, D., Sassoli de Bianchi, M., Sozzo, S. \& Veloz, T. (2018). On the conceptuality interpretation of quantum and relativity theories. 
{\it Foundations of Science}. Online First, doi: 10.1007/s10699-018-9557-z.

\bibitem{melucci2015} Melucci, M. (2015). \emph{Introduction to Information Retrieval and Quantum Mechanics}, The Information Retrieval Series, 35. Berlin Heidelberg: Springer-Verlag, doi: 10.1007/978-3-662-48313-8.

\bibitem{Aerts2005} Aerts, D. (2005). Towards a new democracy: Consensus through quantum parliament. In D. Aerts, B. D'Hooghe and N. Note (Eds.), {\it Worldviews, Science and Us, Redemarcating Knowledge and its Social and Ethical Implications (pp. 189--202)}. Singapore: World Scientific.

\bibitem{Feynman1964} Feynman, R. P. , Leighton, R. B. \& Sands, M. (1964), {\it The Feynman Lectures on Physics Volume III}, New York: Addison-Wesley (revised and extended edition in 2005).

\bibitem{aerts2014} Aerts, D. (2014). Quantum theory and human perception of the macro-world. {\it Frontiers in Psychology 5}, 554, doi: 10.3389/fpsyg.2014.00554.

\bibitem{AertsSassoli2017} Aerts, D. \& Sassoli de Bianchi, M. (2017a). \emph{Universal Measurements}. Singapore: World Scientific.

\bibitem{spinwind01} Aerts, D., Arg\"uelles, J., Beltran, L., Geriente, S., Sassoli de Bianchi, M., Sozzo, S. \& Veloz, T.  (2018). Spin and Wind Directions I: Identifying Entanglement in Nature and Cognition. 
{\it Foundations of Science 23}, pp. 323--335, doi: 10.1007/s10699-017-9528-9.

\bibitem{spinwind02} Aerts, D., Arg\"uelles, J., Beltran, L., Geriente, S., Sassoli de Bianchi, M., Sozzo, S. \& Veloz, T. (2018). Spin and Wind Directions II: A Bell State Quantum Model. {\it Foundations of Science 23}, pp. 337--365, doi: 10.1007/s10699-017-9530-2.

\bibitem{a2009a} Aerts, D. (2009). Quantum structure in cognition. {\it Journal of Mathematical Psychology 53}, pp. 314--348.

\bibitem{h1988a} Hampton, J. A. (1988a). Overextension of conjunctive concepts: Evidence for a unitary model for concept typicality and class inclusion. {\it Journal of Experimental Psychology: Learning, Memory, and Cognition 14}, pp. 12--32.

\bibitem{h1988b} Hampton, J. A. (1988b). Disjunction of natural concepts. {\it Memory \& Cognition 16}, 579--59.

\bibitem{TCS2017} Aerts, D., Arg\"uelles, J., Beltran, L., Beltran, L., Geriente, S., Sassoli de Bianchi, M., Sozzo, S. \& Veloz, T. (2018). Towards a Quantum World Wide Web. {\it Theoretical Computer Science}. Online First, doi: 10.1016/j.tcs.2018.03.019.

\bibitem{asv2015} Aerts, D., Sozzo, S. \& Veloz, T. (2015). Quantum structure in cognition and the foundations of human reasoning. {\it International Journal of Theoretical Physics 54}, pp. 4557--4569.

\bibitem{asv2015b} Aerts, D., Sozzo, S., \& Veloz, T. (2015). Quantum structure of negation and conjunction in human thought. \emph{Frontiers in Psychology 6}, 1447, doi: 10.3389/fpsyg.2015.01447.

\bibitem{as2017} Aerts, D., \& Sozzo, S. (2016). Quantum structure in cognition: Origins, developments, successes and expectations. In (Eds.), E. Haven \& A. Khrennikov, {\it The Palgrave Handbook of Quantum Models in Social Science: Applications and Grand Challenges, pp. 157--193}. London: Palgrave \& Macmillan.

\bibitem{Bachetal2013} Bach, R., Pope, D., Liou, S.-H. \& Batelaan, H. (2013). Controlled double-slit electron diffraction. {\it New Journal of  Physics 15}, 033018.

\bibitem{Melucci2012} Melucci, M. (2012). Contextual search: A computational framework. {\it Foundation and Trends in Information Retrieval 6}, Now Publishers, pp.  257--405.

\bibitem{asdb2015b} Aerts, D.  \& Sassoli de Bianchi, M. (2015). The unreasonable success of quantum probability I: Quantum measurements as uniform measurements. {\it Journal of Mathematical Psychology 67}, pp. 51--75.

\bibitem{AertsSassolideBianchi2014} Aerts, D. \& Sassoli de Bianchi, M. (2014). The Extended Bloch Representation of quantum mechanics and the hidden-measurement solution to the measurement problem. {\it Annals of Physics 351}, pp. 975--102, doi: 10.1016/j.aop.2014.09.020. (See also the Erratum: {\it Annals of Physics 366}, pp. 197--198, doi: 10.1016/j.aop.2016.01.001).

\bibitem{AertsSassoli2016} Aerts, D. \& Sassoli de Bianchi, M. (2016). The Extended Bloch Representation of quantum mechanics. Explaining superposition, interference and entanglement. 
{\it Journal of Mathematical Physics 57}, 122110, doi: 10.1063/1.4973356.

\bibitem{asdbianchi2017} Aerts, D. \& Sassoli de Bianchi, M. (2017). Beyond-quantum modeling of question order effects and response replicability in psychological measurements. 
{\it Journal of Mathematical Psychology 79}, pp. 104--120, doi: 10.1016/j.jmp.2017.03.004.

\bibitem{dariano2004} D'Ariano, G. M., Paris, M. G. A. \& Sacchi, M. F. (2004). 
Quantum tomographic methods. {\it Lecture Notes in Physics 649}, pp. 7--58.

\bibitem{a2011} Aerts, D. (2011). Measuring meaning on the World-Wide Web. In D. Aerts, J. Broekaert, B. D'Hooghe and N. Note (Eds.), {\it Worldviews, Science and Us: Bridging Knowledge and Its Implications for Our Perspectives of the World (pp. 304--313)}. Singapore: World Scientific.

\bibitem{aabssv2016} Aerts, D., Arg\"uelles, J., Beltran, L., Beltran, L., Sassoli de Bianchi, M., Sozzo, S. \& Veloz, T. (2017). Testing quantum models of conjunction fallacy on the World Wide Web. 
{\it International Journal of Theoretical Physics 56}, pp. 3744--3756.

\bibitem{assv2018} Aerts, D., Sassoli de Bianchi, M., Sozzo, S. \& Veloz. T. (2018). Modeling human decision-making: An overview of the Brussels quantum approach. {\it Foundations of Science}. Online First, doi: 10.1007/s10699-018-9559-x.

\bibitem{duretal2017}  D\"ur, W.,  Lamprecht, R. \& Heusler, S. (2017). Towards a quantum internet. {\it European Journal of  Physics 38}, 043001. 

\bibitem{Foskettetal1980} Foskett, D. (1980). Thesaurus. In  A. Kent, H. Lancour, and J. Daily, (Eds), {\it Encyclopaedia of Library and Information Science 30}, pp. 416--462. New York: Marcel Dekker.

\bibitem{Agostietal1991}  Agosti, M., Colotti, R. \& Gradenigo, G. (1991). A two-level hypertext retrieval model for legal data. In {\it SIGIR `91: Proceedings of the 14th Annual International ACM SIGIR Conference on Research and Development in Information Retrieval (pp. 316--325)}. New York, NY, USA.

\bibitem{Agostietal1995} Agosti, M., Crestani, F. \& Melucci, M. (1995). Automatic authoring and construction of hypermedia for Information Retrieval. {\it ACM Multimedia Systems 3}, pp. 15--24.

\bibitem{Sordoni2013}  Sordoni, A., He, J. \& Nie, J.-Y. (2013). Modeling latent topic interactions using quantum interference for information retrieval. In {\it Proceedings of the 22nd ACM International Conference on Information \& Knowledge Management - CIKM `13 (pp. 1197--1200)}. New York, NY, USA: ACM Press. \url{https://doi.org/10.1145/2505515.2507854}.

\end{thebibliography}
\end{document}